\documentclass[sigconf,9pt]{acmart}
\usepackage{amsmath,amsfonts}
\usepackage{algorithmic}
\usepackage{algorithm}
\usepackage{array}
\usepackage{textcomp}
\usepackage{stfloats}
\usepackage{url}
\usepackage{verbatim}
\usepackage{graphicx}
\usepackage{color}
\usepackage{subfig}
\usepackage{gensymb}
\usepackage[export]{adjustbox}
\usepackage{multirow}
\usepackage{cancel}
\usepackage{enumitem}
\usepackage{booktabs}
\usepackage{CJKutf8}
\usepackage{graphicx}
\usepackage{booktabs}
\usepackage{overpic}
\usepackage{balance}
\copyrightyear{2026}
\acmYear{2026}
\setcopyright{acmlicensed}\acmConference[SenSys '26]{The 30th Annual International Conference on Mobile Computing and Networking}{May 11–14, 2026}{Saint-Malo, France}
\acmBooktitle{The 24th ACM/IEEE International Conference on Embedded Artificial Intelligence and Sensing Systems (ACM SenSys '26), May 11–14, 2026, Saint-Malo, France}
\def\systemname{WiRainbow}
\begin{document}
%\settopmatter{printacmref=false} % Removes citation information below abstract
%\renewcommand\footnotetextcopyrightpermission[1]{} % removes footnote with conference information in first column
\pagestyle{plain} % removes running headers
\title{\systemname: Single-Antenna Direction-Aware Wi-Fi Sensing via Dispersion Effect}

\author{Zhaoxin Chang$^{1*}$,
    Shuguang Xiao$^{2,3*}$,
    Fusang Zhang$^{4}$,
    Xujun Ma$^{1}$,
    Badii Jouaber$^{1}$,
    Qingfeng Zhang$^{3\#}$,
    Daqing Zhang$^{1,5\#}$
}

\affiliation{
  \institution{$^{1}$SAMOVAR, Telecom SudParis, Institut Polytechnique de Paris,\\
  ~$^{2}$Shenzhen University of Information Technology,
  ~$^{3}$Southern University of Science and Technology,\\
  ~$^{4}$Beihang University and Institute of Software, Chinese Academy of Sciences, ~$^{5}$Peking University \\
 }
 \country{}
}

%Exploiting Dispersion Effect in Wi-Fi Sensing Using Frequency-Scanning Antennas

\begin{abstract}

Recently, Wi-Fi signals have emerged as a powerful tool for contactless sensing. During the sensing process, obtaining target direction information can provide valuable contextual insights for various applications. Existing direction estimation methods typically rely on antenna arrays, which are costly and complex to deploy in real-world scenarios. In this paper, we present \systemname{}, a novel approach that enables single-antenna-based direction awareness for Wi-Fi sensing by leveraging the dispersion effect of frequency-scanning antennas (FSAs), which can naturally steer Wi-Fi subcarriers toward distinct angles during signal transmission. To address key challenges in antenna design and signal processing, we propose a coupled-resonator-based antenna architecture that significantly expands the narrow Field-of-View inherent in conventional FSAs, improving sensing coverage. Additionally, we develop a sensing signal-to-noise-ratio-based signal processing framework that reliably estimates target direction in multipath-rich environments. We prototype \systemname{} and evaluate its performance through benchmark experiments and real-world case studies, demonstrating its ability to achieve accurate, robust, and cost-effective direction awareness for diverse Wi-Fi sensing applications.
\end{abstract}



\thanks{$^*$Co-primary authors.\\$^\#$Corresponding authors.}



\maketitle

\vspace{-0em}
\section{Introduction}

In recent years, Wi-Fi sensing has attracted growing attention due to the widespread deployment of Wi-Fi infrastructure. This ubiquity enables a wide range of applications, including presence detection~\cite{zhou2013towards,wu2015non,zhu2017r,xin2018freesense}, vital sign monitoring~\cite{liu2014wi,wang2016human,wang2017phasebeat,zeng2019farsense}, gesture recognition~\cite{abdelnasser2015wigest,wang2016device,zheng2019zero,wu2020fingerdraw}, activity recognition~\cite{wang2015understanding,wang2016rt,zhang2019towards,hu2021defall}, and tracking~\cite{adib2013see,li2016dynamic,li2017indotrack,qian2018widar2}. While most existing Wi-Fi sensing systems focus on detecting or monitoring human activities, many real-world scenarios also demand awareness of where these activities occur. A promising way to achieve such spatial awareness is to incorporate direction information into sensing, providing valuable context for applications such as smart home automation. For instance, by mapping directions to physical regions in a room, direction-aware sensing enables the system to infer which area a person is moving through. Moreover, direction awareness helps address long-standing challenges in multi-target sensing and interference mitigation. Therefore, incorporating direction information into Wi-Fi sensing systems can enhance their ability to interpret human motion in space, making them more intelligent and adaptive to real-world environments.

Traditionally, achieving direction awareness for sensing relies on antenna arrays~\cite{xiong2013arraytrack,kumar2014accurate,wang2022wi}, either for Angle-of-Arrival (AoA) estimation at the receiver or beam steering at the transmitter. While various commodity Wi-Fi devices are equipped with multi-antenna arrays for MIMO, this architecture is tailored for communication. Specifically, MIMO relies on feedback-based channel sounding with user devices, followed by beamforming signals toward them~\cite{wu2023enabling,yi2024bfmsense,xu2024beamforming}. Extending this framework to sense passive, non-cooperative targets would require modifications to the protocol stack. For example, VersaBeam~\cite{he2025versabeam} has demonstrated that communication and sensing inherently compete for beamforming resources, so optimizing one inevitably degrades the other. This implies that in MIMO-based Wi-Fi systems, communication and sensing must be jointly optimized through complex mechanisms to achieve a balance. Furthermore, for commodity Wi-Fi devices, the number of antennas may be insufficient for accurate angle resolution, or their placement may not be optimized for AoA estimation~\cite{liu2020survey,korogodin2018impact}. Rather than relying on the existing MIMO architecture for angle estimation and the following sensing tasks, we ask a fundamental question: \textit{\textbf{Can direction awareness in Wi-Fi sensing be achieved using only a single antenna?}} In this paper, we propose a novel design that enables direction-aware sensing using a single, sensing-dedicated antenna, without relying on the MIMO communication architecture or antenna arrays.

\begin{figure}[!t]
	\includegraphics[height=1.8in]{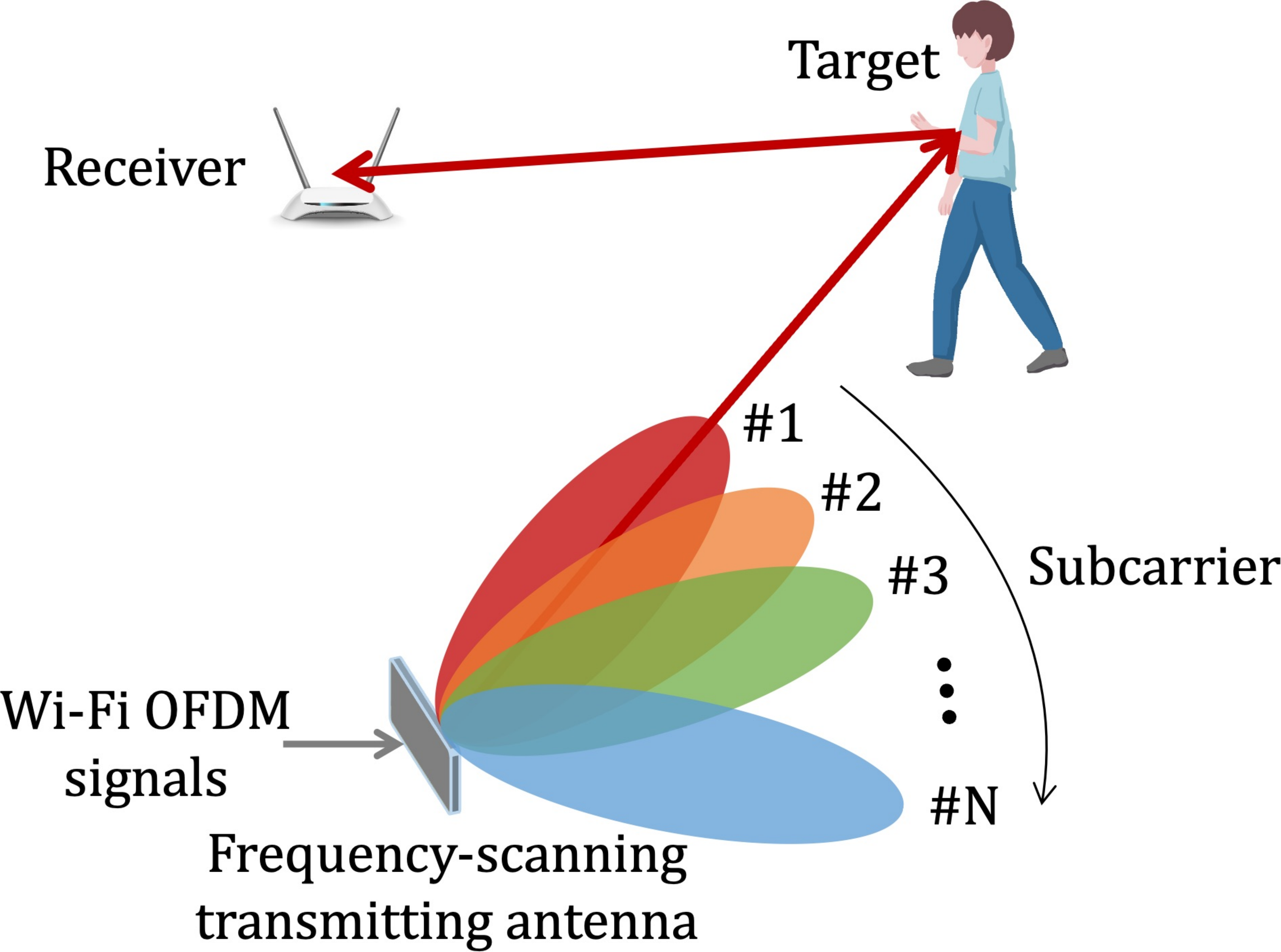}
	\centering{}
	\caption{Using FSAs for direction-aware Wi-Fi sensing.}
	\label{fig1_1}
        \vspace{-0em}
\end{figure}

To achieve this objective, we propose leveraging the inherent frequency-dependent directional dispersion effect of frequency-scanning antennas (FSAs). Unlike conventional omnidirectional or directional antennas, which radiate energy uniformly or similarly across frequencies, FSAs emit different frequency components toward distinct angles, effectively forming frequency-dependent directional beams. This phenomenon is analogous to how a prism disperses white light into different colors. FSAs have been successfully applied in direction estimation for active signal sources across various wireless systems, including RFID~\cite{martinez2011frequency,gil2021frequency,gil2022direction}, Bluetooth~\cite{poveda2020frequency}, Wi-Fi~\cite{sun2023bifrost}, and THz~\cite{ghasempour2020single,kludze2022quasi}. Notably, Wi-Fi adopts Orthogonal Frequency-Division Multiplexing (OFDM), which naturally comprises multiple subcarriers at different frequencies. When an FSA is used for signal transmitting, each subcarrier is inherently radiated toward a distinct spatial direction. Figure~\ref{fig1_1} illustrates the use of FSAs for direction-aware Wi-Fi sensing. Intuitively, when a human target moves within the antenna’s field-of-view (FoV), subcarriers aligned with the target’s direction exhibit more significant signal variations due to human-induced fluctuations. Therefore, by identifying the one subcarrier most affected, the target angle can be estimated. This property also allows selective extraction of the signal corresponding to specific directions,  bringing especially benefits in scenarios involving multiple subjects or dynamic interference. Together, these functionalities underscore the core advantage of the envisioned design, which has potential for lightweight and effective direction awareness in Wi-Fi sensing.

While this concept is theoretically promising, we quickly identify two critical challenges that limit its effectiveness in practical scenarios. First, conventional FSAs, such as traveling-wave antennas, offer a highly limited FoV within the narrow bandwidth typically available for Wi-Fi. For instance, in the 5 GHz band, a 160 MHz bandwidth yields just a 22$\degree$ FoV~\cite{sun2023bifrost}, which is far too restrictive for practical sensing applications. This limitation arises because traditional traveling-wave antenna designs induce minimal angle dispersion effect across closely spaced frequencies, thereby restricting the beam scanning range and sensing coverage. The second challenge arises when we try to achieve target direction estimation. Note that most existing FSA-based systems focus on estimating the directions of active signal sources, typically by measuring signal strength across frequencies. However, human sensing relies on capturing weak reflections from moving targets. In rich multipath indoor environments, strong static reflections can occur at various angles, making it difficult to distinguish the true direction of the target based solely on signal strength. Therefore, a robust method is needed to accurately identify the subcarrier with the strongest target reflection in complex multipath indoor environments for reliable direction estimation. Addressing these challenges is crucial for enabling reliable and effective direction awareness using FSAs in Wi-Fi sensing applications.

To this end, this paper presents \systemname{}, an integrated hardware-software solution that enables direction awareness for Wi-Fi sensing using a single-antenna configuration. \systemname{} introduces two key innovations:

\textbf{Antenna design.} We reveal that the fundamental limitation of traditional FSA designs lies in the use of traveling-wave structures, which induce insufficient phase dispersion across the narrow frequency band of Wi-Fi. To overcome this, we propose a novel antenna design that incorporates resonators as the radiating elements. By leveraging electromagnetic resonance and inter-resonator coupling effects, our design greatly amplifies the sensitivity of phase dispersion to frequency variations. As a result, even small frequency differences can produce substantial angle dispersion, significantly expanding the antenna’s FoV. Specifically, the proposed antenna achieves a 60$\degree$ FoV at the 5 GHz band, which is nearly triple that of existing FSA-based designs.

\textbf{Signal processing.} We also develop a robust signal processing framework that quantifies motion-induced signal variations across Wi-Fi subcarriers. By reliably estimating the Sensing-Signal-to-Noise Ratio~(SSNR) of each subcarrier, our method enables accurate direction estimation even in multipath-rich indoor environments. Furthermore, the algorithm is designed to adapt to different motion scales and speeds, offering a versatile and resilient solution applicable to a wide range of Wi-Fi sensing scenarios.

To validate the effectiveness of \systemname{}, we prototype the proposed design and conduct extensive evaluations. Results confirm the effectiveness and robustness of \systemname{} across diverse sensing scenarios. We showcase the its use in direction-aware walking area detection and interference-resistant multi-target respiration monitoring. The main contributions of this paper are as follows:
\begin{itemize}[leftmargin=*]
\item We propose and validate the feasibility of FSAs for direction-aware Wi-Fi sensing. To the best of our knowledge, this is the first work that enables direction awareness using only a single antenna.
\item We present a novel FSA architecture that substantially expands the antenna’s FoV. Moreover, we introduce a dedicated signal processing framework that reliably identifies human-reflected signals among multipath interference for angle estimation.
\item We prototype the system and conduct both benchmark and real-world evaluations. Experiment results demonstrate robust direction estimation capability, as well as strong performance in multi-target and interference-resistant sensing across diverse indoor environments.
\end{itemize}

\section{Preliminary}

In this section, we first introduce the basis of Wi-Fi sensing. Then, we briefly present the concept and principle of the dispersion effect of FSAs.

\begin{figure}[!t]
        \centering
        \vspace{-0.0em}
        \subfloat[Wi-Fi sensing scenario]{
         \includegraphics[height=0.95in]{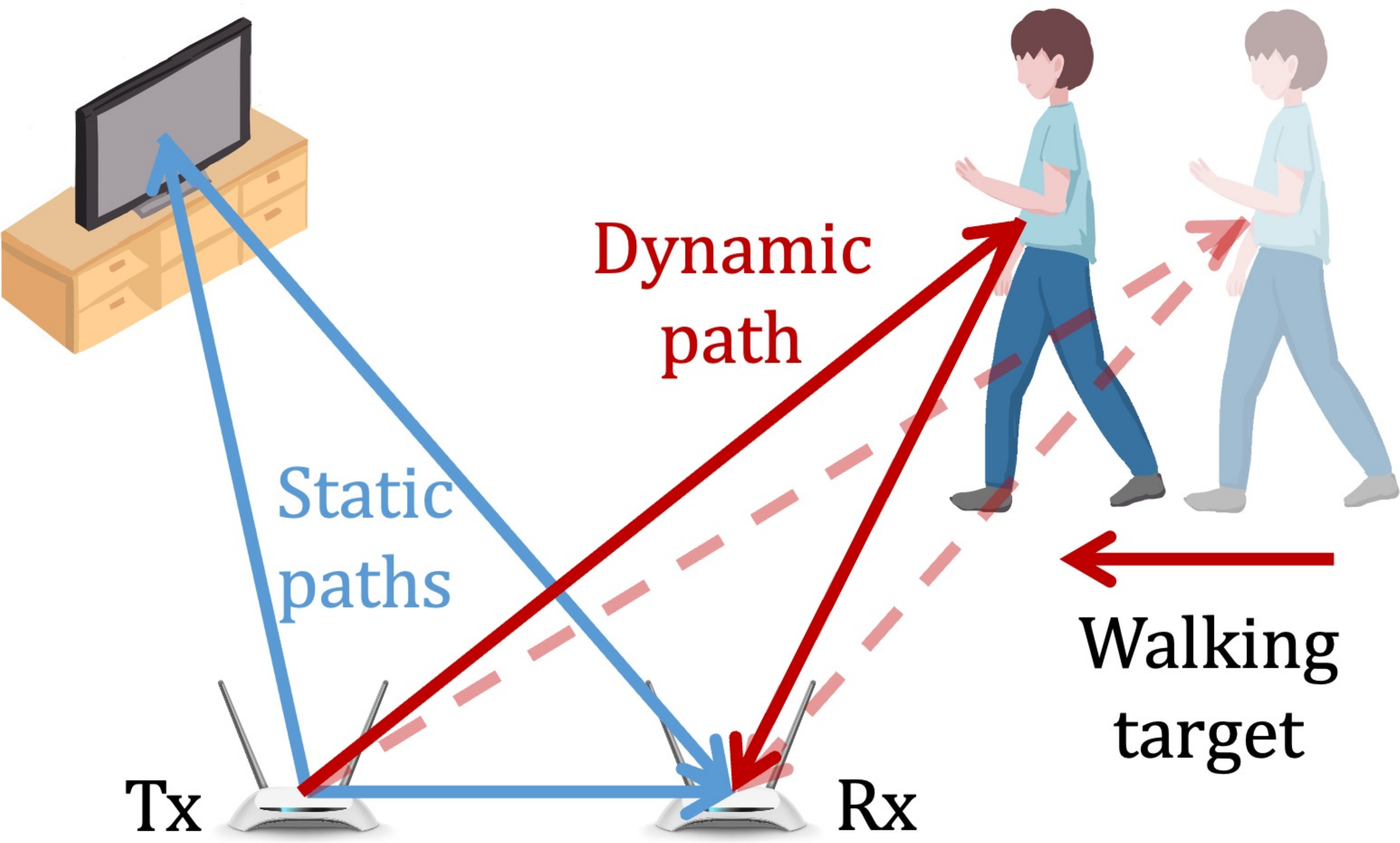}
            \label{fig2_1a}
        }
        \hspace{0.0in}
        \subfloat[CSI on the I-Q plane]{
         \includegraphics[height=0.95in]{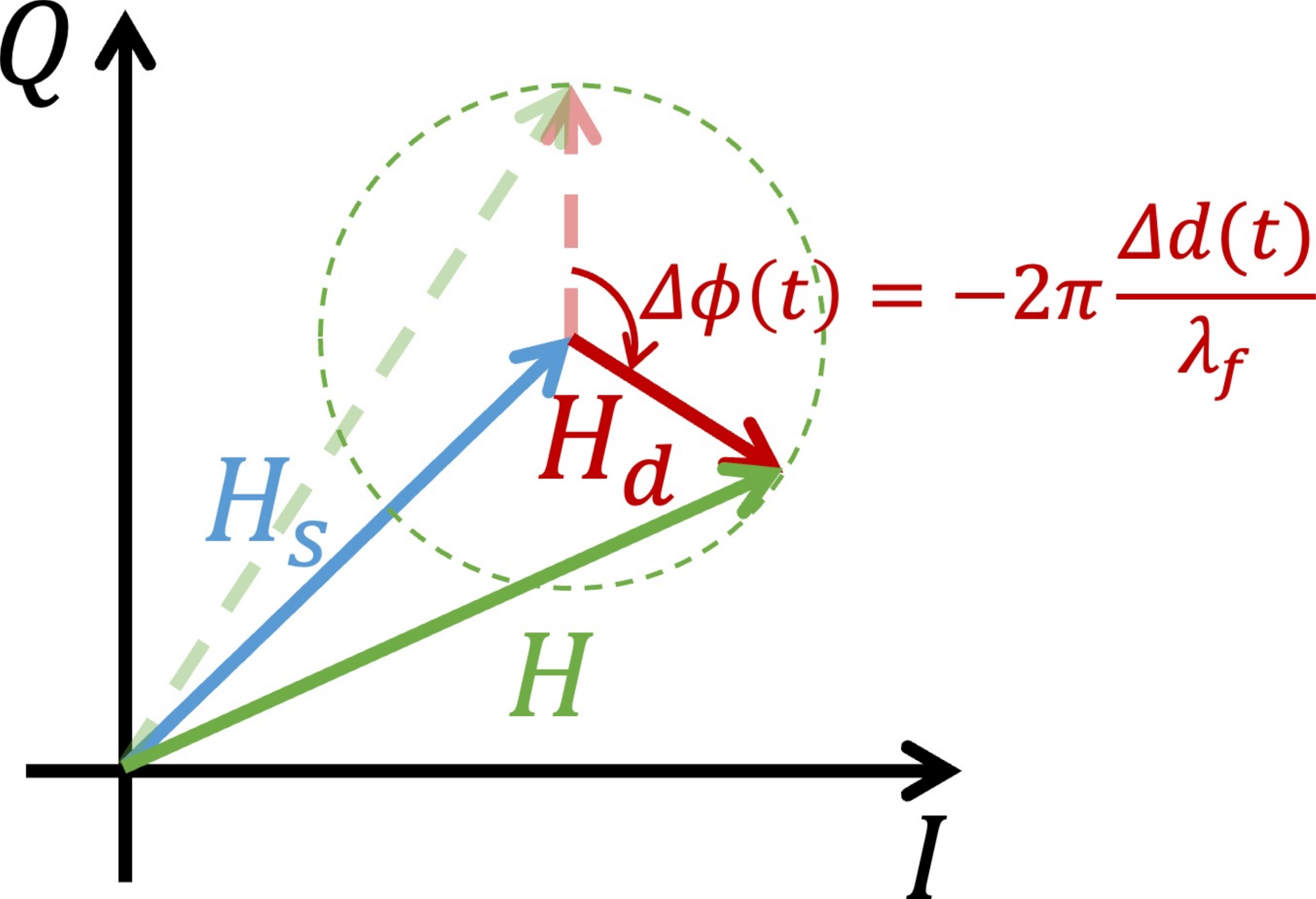}
            \label{fig2_1b}
        }
        \vspace{-0.0in}
        \caption{Principle of Wi-Fi sensing.}
        \vspace{-0.0em}
        \label{fig2_1}
        %\vspace{-0.15in}
\end{figure}

\subsection{Wi-Fi Sensing Primer}\label{sec21}

As shown in Figure~\ref{fig2_1a}, signals emitted by a transmitter (Tx) propagate through multiple paths before arriving at the receiver (Rx). These paths include the Line-of-Sight (LoS), reflections from static objects (static paths), and reflection from the moving target (dynamic path). As a result, the received signal is a combination of the static component ($H_s(f)$) and dynamic component ($H_d(f,t)$). Typically, Channel State Information (CSI) is utilized to characterize Wi-Fi signal propagation, which can be expressed as:
\begin{equation}\label{eq2_1}
\begin{split}
H(f,t) &= H_s(f) + H_d(f,t) + n(f,t) \\
&= H_s(f) + A(f,t)e^{-j2\pi \frac{d(t)}{\lambda_f}} + n(f,t),
\end{split}
\end{equation}
where $n(f,t)$ represents noise, $A(f,t)$ is the amplitude of the dynamic component, $d(t)$ is the dynamic path length, and $\lambda_f$ denotes the wavelength. Notably, when the dynamic path length changes, the CSI rotates on the I-Q plane as illustrated in Figure~\ref{fig2_1b}. Consequently, the amplitude and phase of the CSI change over time. By extracting these variations, human motion features such as displacement and Doppler speed can be estimated for sensing. Note that due to the lack of synchronization between Wi-Fi transceivers, the raw CSI extracted from the receiver often suffers from random phase offsets, complicating accurate phase analysis. To address this issue, CSI ratio approach~\cite{zeng2019farsense} can be employed, which effectively removes phase offsets by taking the ratio of CSI from two receiving antennas sharing the same offsets. In this paper, we utilize CSI ratio to deal with the phase offsets before signal processing. Equation~\ref{eq2_1} is the CSI after offset cancellation.

\vspace{-0em}
\subsection{Basic Principle of FSAs} \label{sec22}

\begin{figure}[!t]
\includegraphics[height=1.7in]{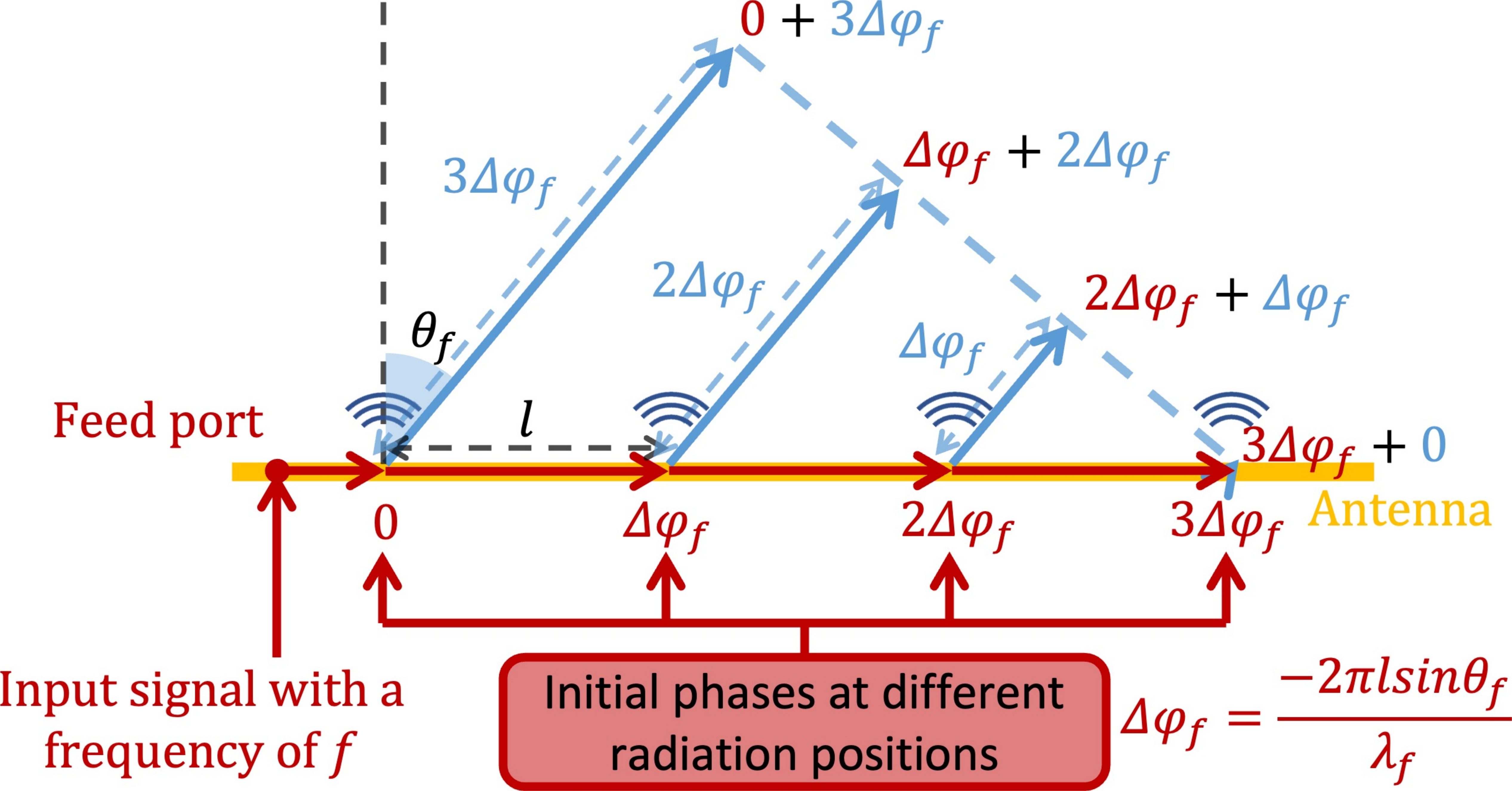}
	\centering{}
	\caption{Principle of frequency-scanning antennas.}
    \vspace{-0.0em}
	\label{fig2_2}
    \vspace{-0.0em}
\end{figure}

In this section, we briefly introduce the fundamental concept behind the frequency-scanning property of FSAs. Unlike traditional antennas, FSAs enable signals to propagate progressively along their structures while simultaneously radiating energy into space. A typical implementation is the traveling-wave antenna, such as the leaky-wave antenna (LWA), which consists of multiple periodically arranged radiating elements. As illustrated in Figure~\ref{fig2_2}, when a signal of frequency $f$ is fed into the antenna, it travels along the structure and ``leaks'' energy into the environment at several positions (e.g., four radiation points in the figure). Because each segment of the signal propagates a certain distance before radiating, it accumulates a distinct phase shift. Specifically, after traveling a distance $l$ between adjacent elements, the signal experiences a phase delay of $\Delta \varphi_f$. As a result, the signals emitted from the four elements exhibit progressive initial phase values of 0, $\Delta \varphi_f$, $2\Delta \varphi_f$, and $3\Delta \varphi_f$, respectively. These radiated signals combine in space, and due to their phase progression, they constructively interfere in a specific direction where all components align in phase. This forms a strong, directional beam at frequency $f$. This unique beam direction at a frequency of $f$ can be expressed as:
\vspace{-0em}
\begin{equation}\label{eq2_2}
    \theta_f = \arcsin \frac{-\lambda_f \Delta \varphi_f}{2\pi l}.
\end{equation}
It is important to note that as signals propagate along the antenna, different frequencies naturally experience distinct phase delays ($\Delta \varphi_f$). As a result of these frequency-dependent phase delays, signals at different frequencies are steered toward different spatial directions according to Equation~\ref{eq2_2}, which is the fundamental principle behind FSA-based beam-scanning.

This principle is analogous to conventional beamforming techniques in antenna arrays, where beam directions are steered by actively introducing phase differences between multiple antenna elements. In contrast, FSAs achieve a similar effect in a more elegant and passive manner by leveraging the intrinsic propagation of signals along the antenna structure to induce progressive phase shifts. This built-in phase dispersion eliminates the need for external phase shifters or control circuitry, thereby significantly simplifying the system architecture.

%This principle resembles that of conventional beamforming technique using antenna arrays, which steer beam directions by actively adjusting the phase difference between individual antenna elements. In contrast, FSAs achieve this effect more elegantly by exploiting the intrinsic propagation of signals along their structure to induce the necessary phase differences. This passive phase dispersion mechanism significantly simplifies the system architecture.

\subsection{Challenges of Utilizing FSAs for Sensing}\label{sec23}

In Wi-Fi sensing, acquiring target direction information is valuable, as it provides important contextual awareness for various applications. FSAs present an alternative to multi-antenna configurations by enabling direction awareness with minimal hardware. Since Wi-Fi adopts OFDM modulation, subcarriers at different frequencies are transmitted simultaneously. When an FSA is used as the transmitting antenna, each subcarrier is emitted to a distinct direction due to the antenna’s inherent frequency-dependent beam-steering property. Intuitively, the subcarrier aligned with the target’s direction is most affected by human-induced signal fluctuations, allowing us to infer the target’s direction by analyzing the received Wi-Fi signals or extracting the signals from desired directions.

However, we quickly identify two key challenges in applying existing FSA designs to Wi-Fi sensing, arising from both antenna design and signal processing perspectives. First, conventional FSAs typically adopt a traveling-wave structure, in which signals propagate linearly along the antenna. The angular separation between subcarriers, i.e., the difference in their beam directions, depends on their frequency spacing. However, Wi-Fi operates over a relatively narrow frequency band (usually tens to hundreds of megahertz), resulting in minimal differences in propagation phase delay ($\Delta \varphi_f$) across subcarriers. Consequently, the beams associated with different subcarriers are poorly separated in angle. Additionally, as signals propagate along the antenna, they experience only negative phase delays. This restricts beam steering to a single side relative to the antenna’s forward direction, as illustrated in Figure~\ref{fig2_2}. Together, these factors lead to a severely limited FoV when using FSAs within the Wi-Fi band, which is often far narrower than what is needed for practical indoor sensing. For example, a typical traveling-wave-based FSA achieves a FoV of only 22$\degree$ across a 160 MHz bandwidth at 5 GHz~\cite{sun2023bifrost}. Therefore, \textit{\textbf{the extremely limited FoV presents a fundamental challenge in antenna design.}}

Second, when attempting to estimate the target’s direction, conventional FSAs face another major challenge on the signal processing side. Traditional designs determine the signal source direction by comparing received signal strengths (e.g., SNRs) across different frequencies. However, this straightforward approach is insufficient for sensing, particularly in indoor environments. Severe multipath effects cause reflections from static objects, in addition to the intended human target, to produce strong signal responses at certain frequencies corresponding to their directions. As a result, relying solely on signal strength makes it difficult to accurately identify the direction of the target. Hence, \textbf{\textit{the lack of an effective method to estimate target direction in  multipath-rich environments poses a challenge in signal processing.}}

In the following sections, we propose novel antenna design and signal processing methods to address these two critical challenges, enabling direction awareness using FSAs for Wi-Fi sensing.

\section{Methodology}
\label{sec3}

\begin{figure}[!b]
\vspace{-0em}
\includegraphics[height=1.00in]{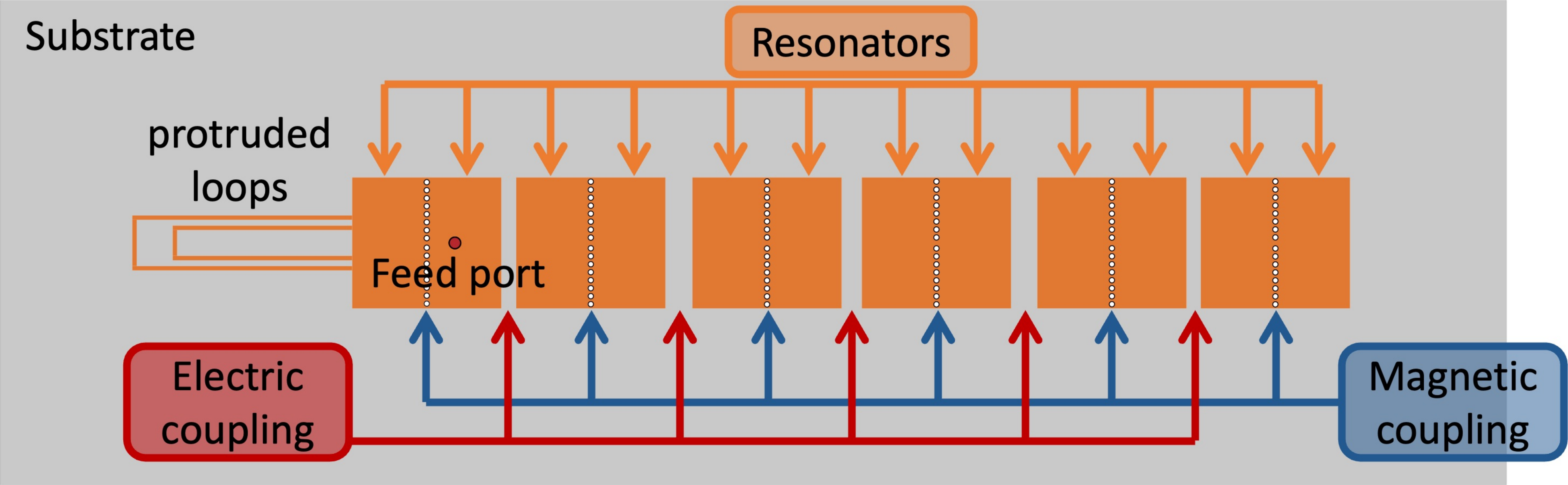}
	\centering{}
        \vspace{-0em}
	\caption{Illustration of the proposed antenna design.}
	\label{fig3_1_0}
\end{figure}

\subsection{Antenna Design}

As introduced in Section~\ref{sec23}, existing FSA designs often suffer from a limited FoV, which restricts their effectiveness in practical sensing applications. This limitation arises from the small phase delay ($\Delta \varphi_f$) differences between closely spaced frequencies within the narrow Wi-Fi band, leading to insufficient frequency dispersion. To overcome this issue, we propose a novel antenna design that enhances frequency-dependent dispersion by increasing the phase delay variation across subcarriers. Our approach leverages two fundamental electromagnetic principles: \textbf{\textit{resonance}} and \textbf{\textit{coupling}}. Specifically, we introduce a coupled-resonator architecture, as illustrated in Figure~\ref{fig3_1_0}. The antenna comprises 12 resonator elements fabricated on a double-layer printed circuit board (PCB). During operation, Wi-Fi signals enter through a feed port and sequentially propagate through resonators. Each resonator acts as an individual radiator, emitting signals into the surrounding environment. Signal propagation between resonators is enabled by two types of coupling mechanisms: electric coupling, achieved by a small gap between adjacent resonators, and magnetic coupling, realized using small conductive vias. As Wi-Fi subcarriers propagate within and between resonators, they accumulate frequency-dependent phase delays, resulting in distinct beam directions for different subcarriers. This mechanism enhances the dispersion capability and significantly expands the antenna's FoV. Further details of the resonator structure are provided in Section~\ref{sec311}, and the coupling mechanisms are elaborated in Section~\ref{sec312}.

\begin{figure}[!b]
\includegraphics[height=1.7in]{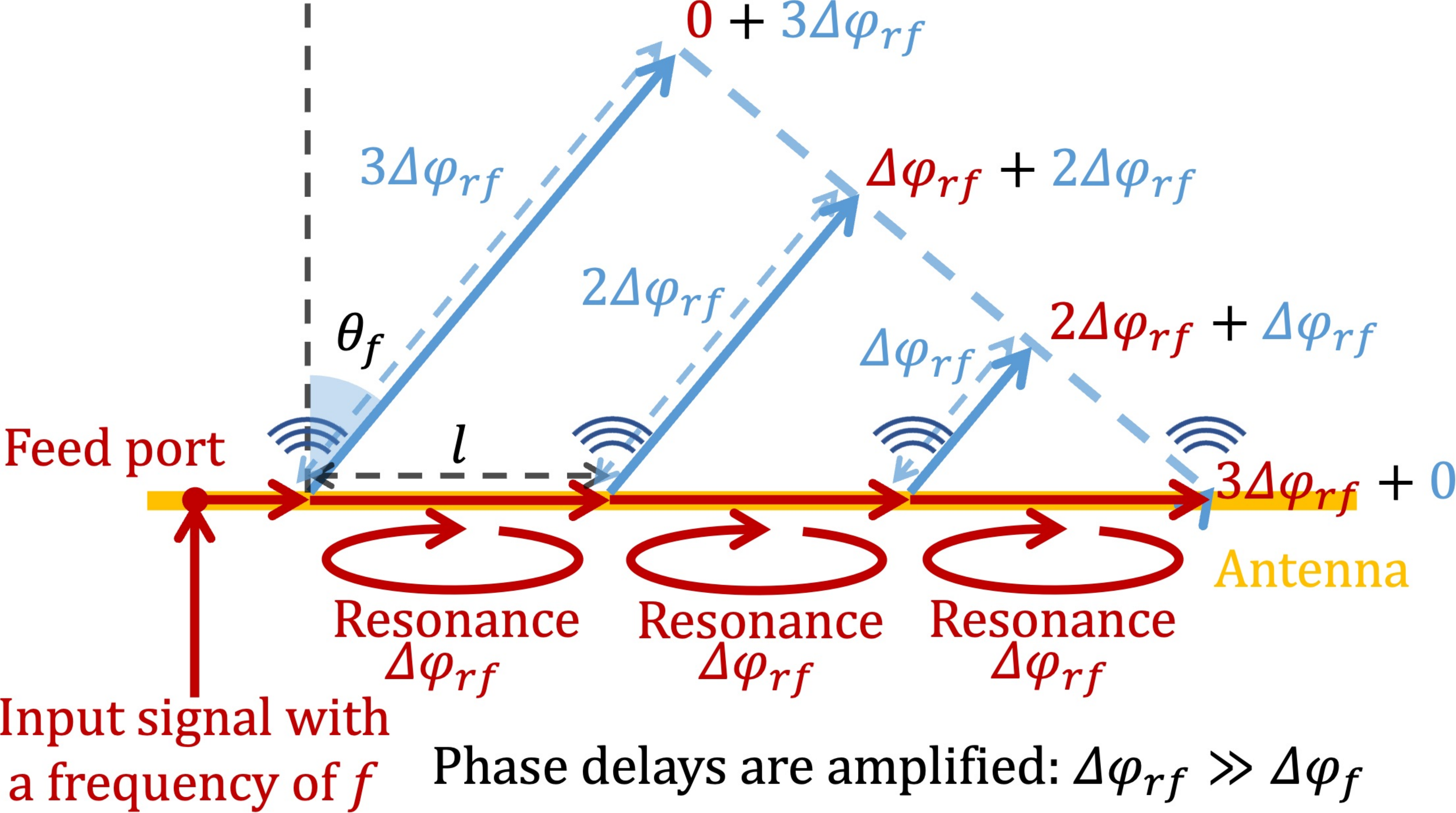}
	\centering{}
	\caption{Principle of resonator-based architecture.}
    \vspace{-0em}
	\label{fig3_1_1}
\end{figure}

\subsubsection{Resonator-based Architecture}\label{sec311}
We propose a resonator-based antenna architecture that significantly enhances phase delay by leveraging the electromagnetic resonance effect. In this design, metal patches are fabricated on the substrate as resonators. Their dimensions (length and width) and the substrate’s material properties (thickness and permittivity) are carefully selected to define the \emph{resonant frequency} $f_0$, at which signal energy becomes strongly confined within the patch.

As shown in Figure~\ref{fig3_1_1}, when the input Wi-Fi signal frequency approaches $f_0$, the signal undergoes resonance within the patch, oscillating repeatedly instead of passing through directly. This resonance effect is similar to an echo chamber, causing the signal to accumulate additional phase delays through repeated internal propagation. Consequently, even a slight difference in signal frequency results in significant variations in phase delay. According to the fundamental FSA principle introduced in Section~\ref{sec22}, greater phase delay differences among frequencies directly translate into larger angular separations between beam directions. This enables a substantially wider FoV than conventional traveling-wave-based FSAs.

\begin{figure}[!t]
\includegraphics[height=1.4in]{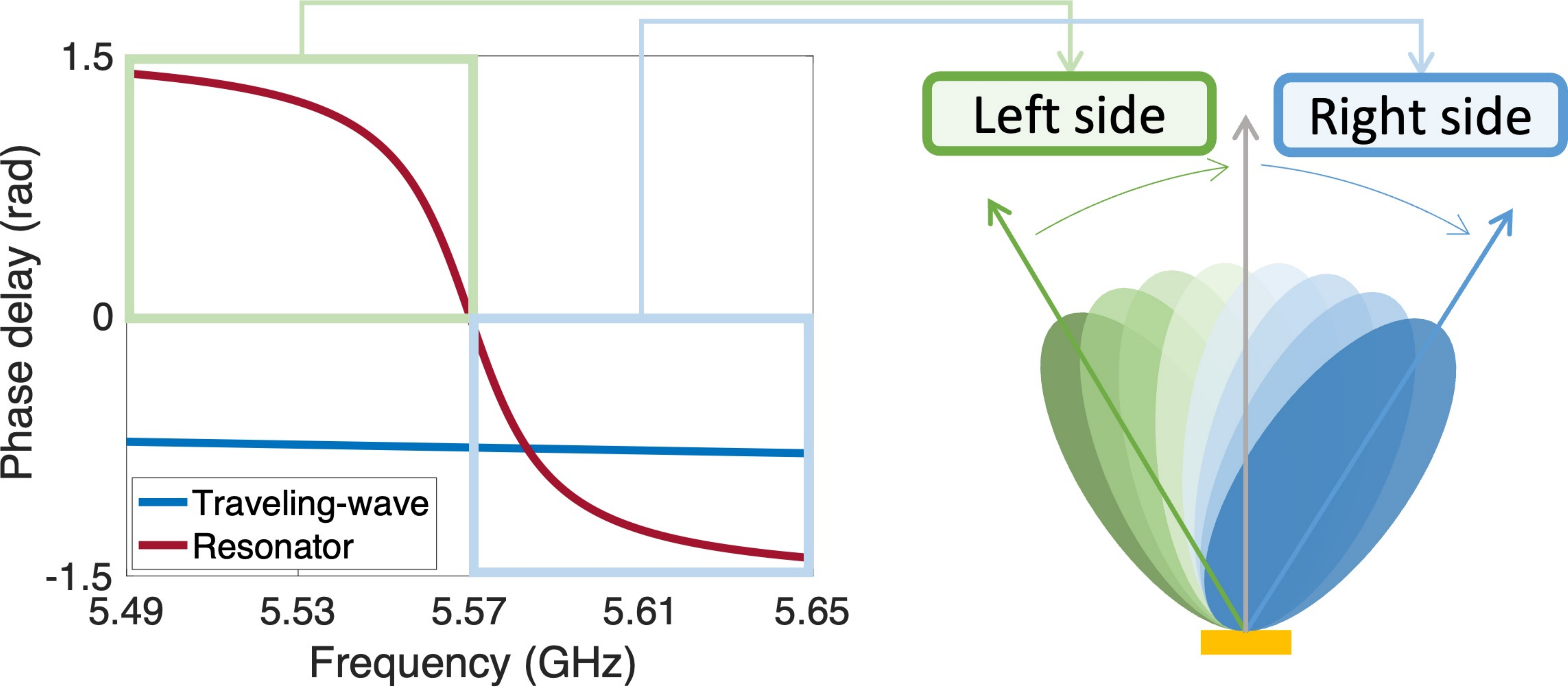}
	\centering{}
	\caption{Frequency-dependent phase delays and the impact on beam steering.}
    \vspace{-0em}
	\label{fig3_1_2}
\end{figure}

Physically, the phase delay introduced by a resonator patch can be calculated as follows:
\begin{equation}\label{eq3_02}
    \Delta \varphi_{rf} = \arctan\{Q(\frac{f_0}{f} - \frac{f}{f_0})\},
\end{equation}
where the quality factor $Q$ quantifies the strength of the resonance. A higher $Q$ value indicates a stronger resonance effect. The value of $Q$ depends primarily on the patch dimensions and substrate material properties. Figure~\ref{fig3_1_2} presents simulation results of $\Delta \varphi_{rf}$, with parameters set as follows: $Q=200$, a resonant frequency $f_0=5.57$ GHz, and input signal frequencies ranging from 5.49 to 5.65 GHz (i.e., Wi-Fi channel 114). The results demonstrate that even small frequency variations can lead to substantial phase delays in the resonator-based design, in contrast to conventional traveling-wave antennas. This enhanced phase dispersion enables greater angular separation between beams formed by different subcarriers. Moreover, the results show that signal frequencies below $f_0$ produce positive phase delays, steering the beam toward negative angles (to the left within the FoV). In contrast, frequencies above $f_0$ yield negative phase delays, directing the beam toward positive angles (to the right). By setting the resonant frequency $f_0$ at the center of the operating band, the antenna achieves a symmetric beam distribution across the FoV. Together, these two properties, i.e., the amplification of frequency-dependent phase delay and the symmetry of beam steering, allow our resonator-based design to achieve a significantly wider field of view than traditional FSA structures.

\subsubsection{Inter-Resonator Coupling Mechanisms}\label{sec312}

While the proposed resonator-based architecture significantly enhances the intra-patch phase delay, we identify another critical challenge. Signals traveling between adjacent resonator patches can experience additional, unintended phase shifts caused by inter-patch coupling. Note that signal transmission between patches relies on the coupling effect, which occurs in two distinct modes depending on the configuration between neighboring patches: electric coupling and magnetic coupling. \textit{\textbf{Electric coupling}} arises when two patches are placed side by side on the substrate, separated by a narrow gap. In this configuration, the time-varying electric field generated by one patch induces electric charges on the adjacent patch, transferring energy primarily through electric field interactions. In contrast, \textit{\textbf{magnetic coupling}} occurs when adjacent patches are connected by conductive vias, which are small metallic holes. In this case, the current in one patch generates a time-varying magnetic field that induces currents in the neighboring patch, thereby transferring energy primarily through magnetic field interactions.

\begin{figure}[!t]
\includegraphics[height=1.35in]{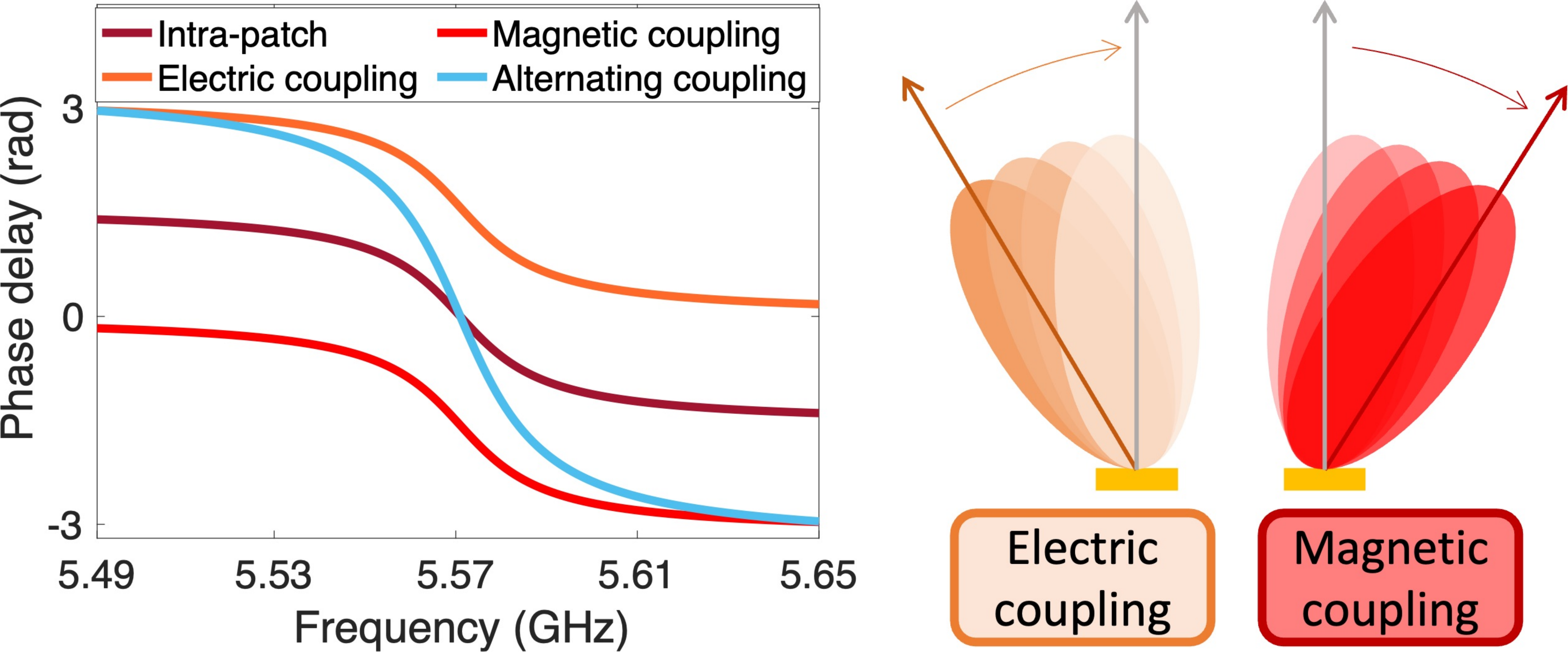}
	\centering{}
    \vspace{-0em}
	\caption{Impact of different types of coupling effects on phase delays.}   
	\label{fig3_1_3}
\end{figure}

Importantly, these two coupling mechanisms have opposite influences on the signal phase~\cite{zhou2022design}. Electric coupling introduces a positive phase delay, whereas magnetic coupling introduces a negative phase delay, steering the beam toward positive angles. Conventional resonator-based antenna designs usually adopt only one of these coupling types, either electric or magnetic. As illustrated in Figure~\ref{fig3_1_3}, when only magnetic coupling is applied, a negative phase delay is added to the inherent intra-patch phase delay. Consequently, the phase delays of all frequencies become negative, and all beams are steered toward the right side of the FoV. Conversely, when only electric coupling is used, the additional positive phase delay causes all beams to shift toward the left side. Therefore, such traditional coupling configurations result in an imbalanced beam distribution that is biased to one side, thereby limiting the overall FoV.

To address this critical limitation, we propose an innovative \textbf{\textit{alternating coupling}} strategy. In this design, adjacent resonator patches are alternately connected using electric and magnetic coupling. For example, magnetic coupling is followed by electric coupling, and this pattern is repeated throughout the array, as illustrated in Figure~\ref{fig3_1_0}. This alternating configuration naturally balances the positive and negative phase delays introduced by the two coupling types, effectively canceling out their cumulative effects. By carefully engineering the antenna structure, we ensure that at the resonant frequency $f_0$, both coupling mechanisms contribute equally. As a result, their opposing phase delays neutralize each other, causing the beam to radiate directly forward, corresponding to 0$\degree$ at the center of the FoV. To further fine-tune this balance, we incorporate two small protruding loops into the first patch to adjust the coupling strength precisely. As shown in Figure~\ref{fig3_1_3}, when the input frequency is lower than $f_0$, electric coupling dominates, resulting in beams being steered toward negative angles. When the frequency is higher than $f_0$, magnetic coupling becomes dominant, causing beams to shift toward positive angles. This design not only expands the FoV but also ensures symmetrical and continuous beam scanning across the entire angular range.

In summary, our proposed antenna architecture addresses two fundamental limitations of traditional FSAs. First, the resonator-based patch design introduces large and frequency-sensitive phase delays, which significantly enhance angular separation between beams and improve the FoV. Second, the alternating coupling strategy mitigates phase imbalance between coupling types, further expanding the FoV while enabling symmetric and smooth beam steering. Together, these innovations result in a broader, more uniform, and highly effective scanning capability, meeting critical demands for practical Wi-Fi sensing applications.

\subsection{Basic Sensing Model}\label{sec32}

In this section, we first establish the FSA-based sensing model. As illustrated in Figure~\ref{fig1_1}, an FSA is used as the transmitting antenna. Let $D(f,\theta)$ denote the frequency-dependent beam pattern, representing the signal strength of frequency $f$ in direction $\theta$. For each frequency $f$, there exists a dominant beam direction $\theta_f$ where the radiated energy is strongest. This beam-steering property leads to a unique signal propagation behavior across subcarriers and modifies the conventional CSI model as follows:
\begin{equation}\label{eq3_1}
\begin{split}
    H(f,t) =& H_s(f) + D(f,\theta_d)H_d(f,t) + n(f,t) 
    \\ =& H_s(f) + D(f,\theta_d)A(f,t)e^{-j2\pi \frac{d(t)}{\lambda_f}} + n(f,t),
\end{split}
\end{equation}
where $\theta_d$ denotes the target’s angle relative to the transmitting FSA. Note that the static channel component $H_s(f)$ differs from that in Equation~\ref{eq2_1} due to the frequency-selective directional radiation of the FSA. In this equation, the subcarrier whose beam direction $\theta_f$ best aligns with $\theta_d$ is expected to capture the strongest dynamic variation from the target. This introduces a new capability for Wi-Fi sensing: subcarriers become spatial filters that are sensitive to different directions. As a result, it becomes possible to selectively extract information from a desired direction while suppressing interference from others. For example, by focusing on the subcarrier corresponding to a specific angle $\theta_f$, the system can observe motion occurring in that direction while minimizing contributions from other directions. This property not only enables direction estimation but also supports robust multi-target sensing and interference mitigation in complex environments.

However, due to multipath effects and the finite beam width of each subcarrier, other subcarriers may also capture noticeable variations. For instance, when the target is close to the FSA, motion-induced signal changes can be observed across a wider subcarrier range. Therefore, if the objective is to estimate target's direction, the key is to identify the subcarrier with the strongest dynamic component. To quantify this, we adopt the SSNR proposed in~\cite{li2022diversense,wang2022placement}, defined as the ratio between the power of the dynamic component and the noise level:
\begin{equation}\label{eq3_2}
\begin{split}
    SSNR = \frac{|H_d(f,t)|^2}{|n(f,t)|^2} = \frac{|D(f,\theta_d)A(f,t)|^2}{|n(f,t)|^2}.
\end{split}
\end{equation}
Here, we assume the target reflection amplitude $A(f,t)$ remains approximately constant across subcarriers. Consequently, the variation in SSNR is primarily driven by $D(f,\theta_d)$, meaning the subcarrier aligned with $\theta_d$ should yield the highest SSNR. In the following section, we present the detailed signal processing approach for SSNR measurement, which enables robust direction estimation.

\subsection{Direction Estimation Based on SSNR Measurement}
To estimate SSNR accurately, it is first necessary to eliminate the static component from the raw CSI, which comprises static, dynamic, and noise terms as shown in Equation~\ref{eq3_1}. Several methods exist for removing the static component, such as mean value subtraction~\cite{li2017indotrack,yu2018qgesture}, tangential direction extraction~\cite{wu2020fingerdraw,zhang2020exploring}, circle fitting~\cite{song2020spirosonic,zhang2019towards}, and Time-domain Difference of CSI (TD-CSI)~\cite{li2024wifi}. In this work, we adopt the TD-CSI approach for its high accuracy and adaptability. The core idea of TD-CSI is that the static component remains nearly constant between consecutive CSI samples, while the dynamic and noise components fluctuate over time. By taking the difference between two adjacent samples, the static part is effectively removed, leaving only the dynamic and noise components. TD-CSI is defined as:
\begin{equation}\label{eq3_3}
\begin{split}
   \Delta H(f,t) &= H(f,t+\Delta t) - H(f,t) \\ &= \Delta H_s(f) + \Delta H_d(f,t) + \Delta n(f,t) \\ 
   &= 2A(f)sin \frac{\Delta \phi(t)}{2}e^{j(\phi(t) + \frac{\Delta \phi(t)}{2} + \frac{\pi}{2})} + \Delta n(f,t),
\end{split}
\end{equation}
where $\Delta t$ is the interval between two CSI samples, $\phi(t) = -2\pi \frac{d(t)}{\lambda_f}$ represents the phase of the dynamic component, $\Delta \phi(t)$ is the phase variation over $\Delta t$, and $\Delta n(f,t)$ is the noise change. During a short time window, the amplitude of the dynamic component can be assumed to remain constant.

\begin{figure}[!b]
\vspace{-0em}
        \centering
        \subfloat[High SSNR]{
        \includegraphics[height=1.18in]{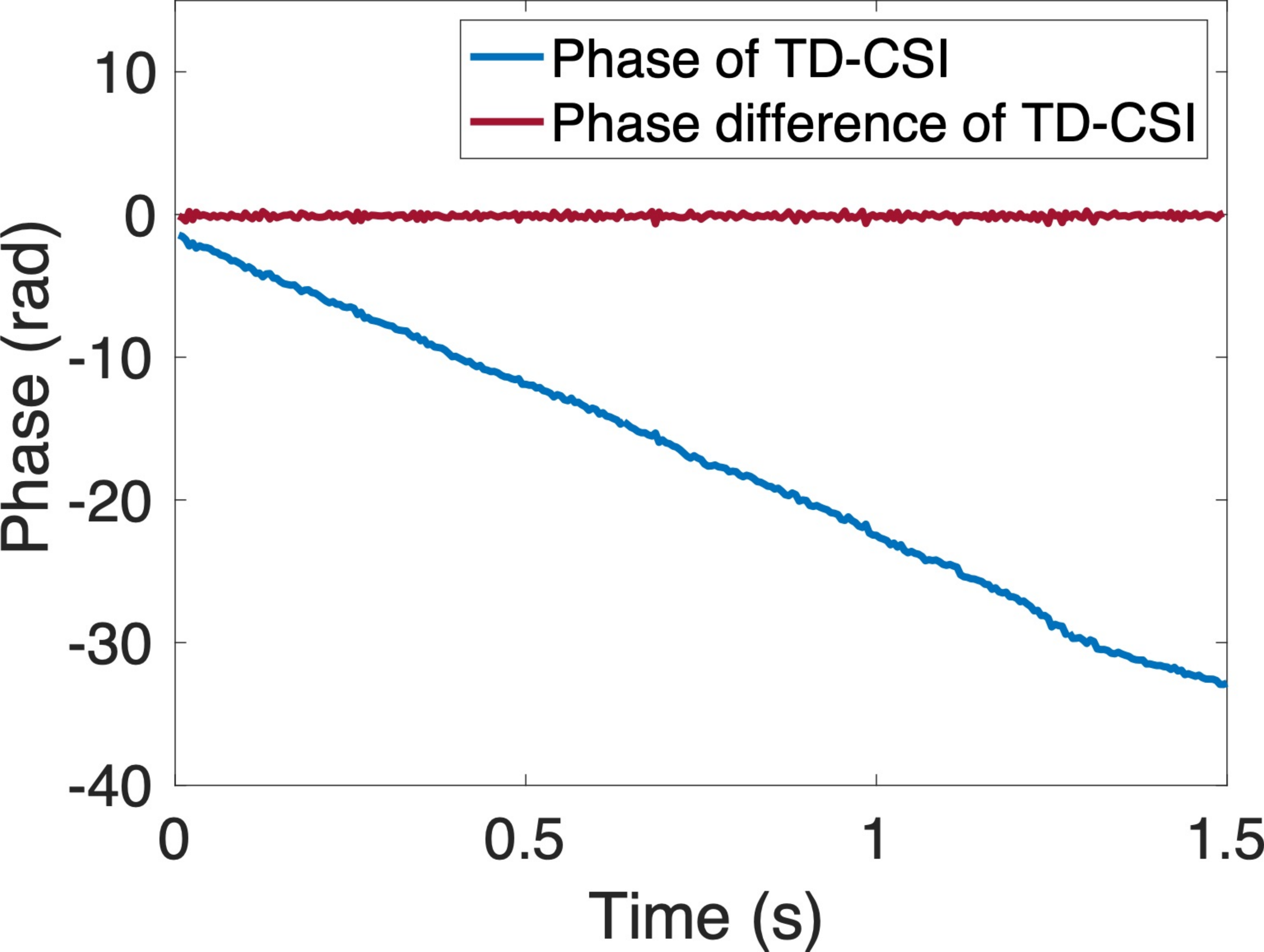}
            \label{fig3_4a}
        }
        \hspace{0.0in}
        \subfloat[Low SSNR]{
        \includegraphics[height=1.18in]{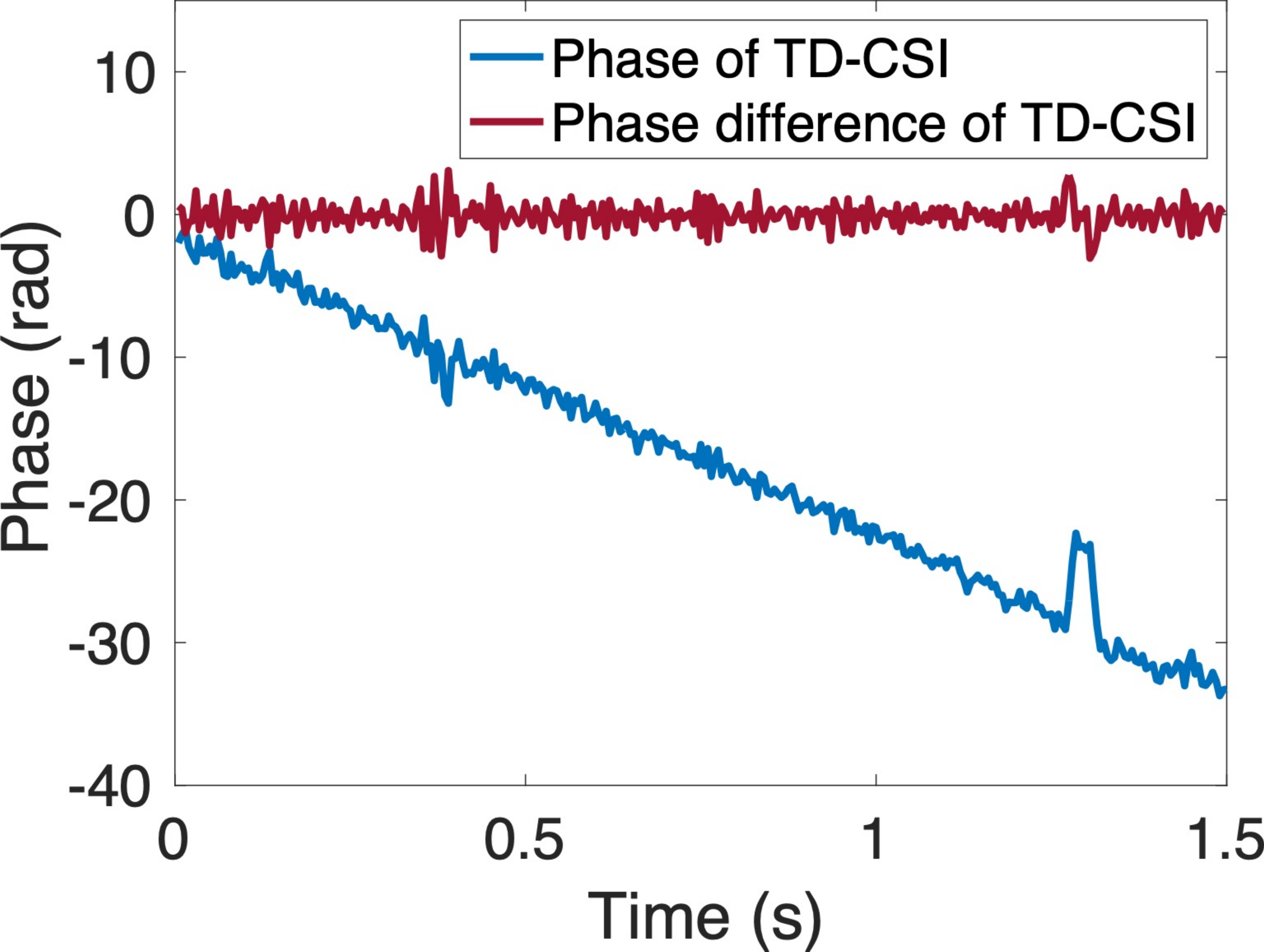}
            \label{fig3_4b}
        }
        \vspace{-0em}
        \caption{Illustration of the phase and phase difference of TD-CSI with high and low SSNR.} 
        \label{fig3_4}
        %\vspace{-0.15in}
\end{figure}

To quantify SSNR from TD-CSI $\Delta H(f,t)$, we leverage the temporal characteristics of the signal. The dynamic component tends to vary smoothly over time due to the physical limitations of human motion speed, while noise fluctuates in an unpredictable manner. Building on this insight, we propose to measure SSNR by analyzing the stability of TD-CSI phase variations. Specifically, when SSNR is high, TD-CSI is dominated by motion-induced signals, leading to smooth phase variations. In contrast, low SSNR results in irregular phase fluctuations dominated by noise. Figure~\ref{fig3_4} illustrates this difference: under high SSNR, the TD-CSI phase progresses smoothly, and the phase difference between consecutive samples remains close to zero. Under low SSNR, the phase varies irregularly, and the differences deviate significantly. We therefore compute the phase differences between consecutive TD-CSI samples and then calculate the variance. A low variance corresponds to stable dynamics and high SSNR, while a high variance indicates noisy signals and low SSNR.

\begin{figure}[!t]
\includegraphics[height=1.3in]{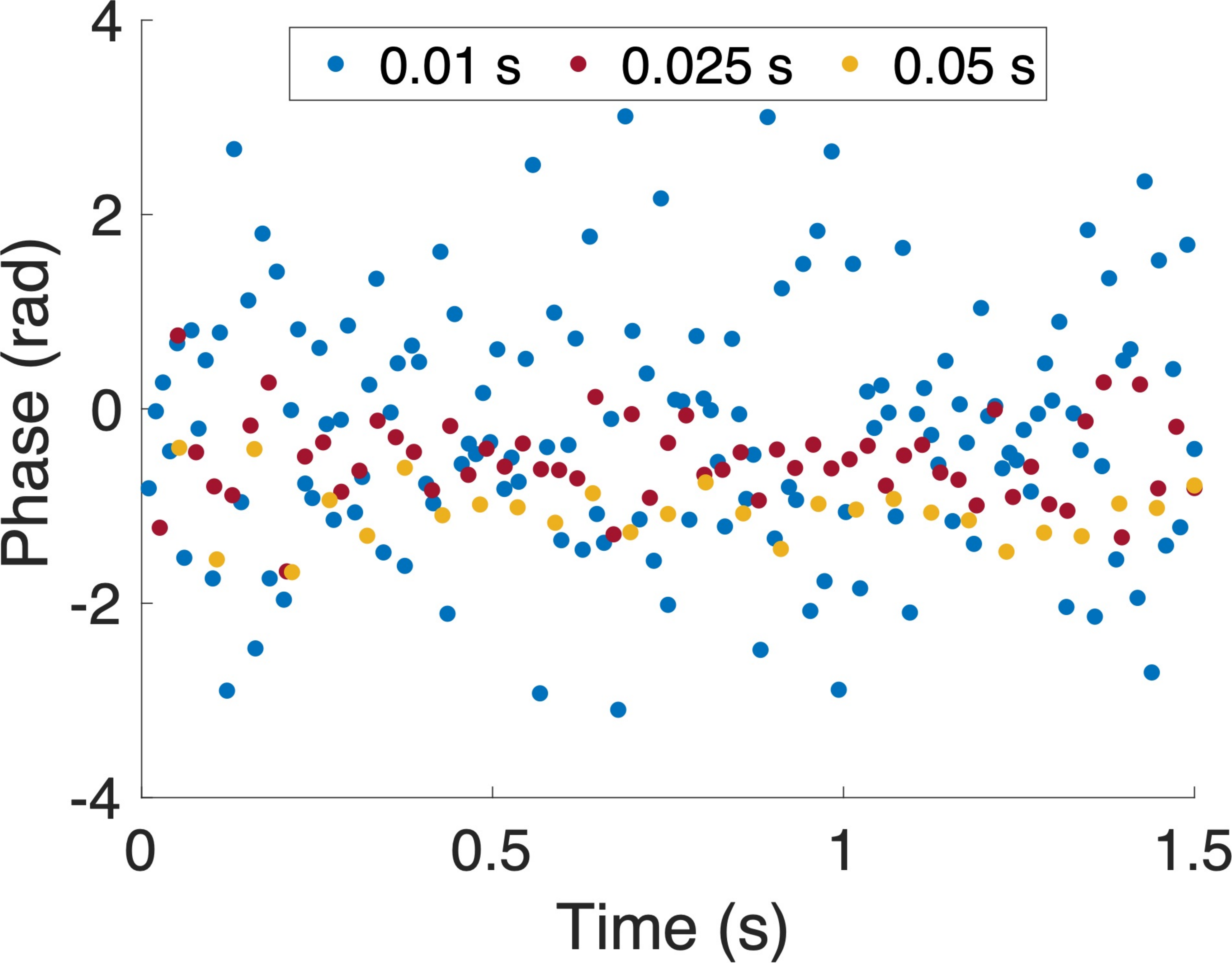}
	\centering{}
    \vspace{-0em}
	\caption{Phase differences of TD-CSIs with different time intervals.}
	\label{fig3_6}
        \vspace{-0em}
\end{figure}

However, during experimental verification, we identify a critical issue: when the target moves at a slow speed, all TD-CSI subcarriers exhibit highly random phase variations. Further analysis reveals that, according to Equation~\ref{eq3_3}, when the target’s speed is low, the phase change of the dynamic component ($\Delta \phi (t)$) remains nearly constant within the time interval ($\Delta t$) used for TD-CSI computation. As a result, even if the SSNR is not particularly low, TD-CSI becomes dominated by noise. To address this issue, an intuitive solution is to increase the time interval for TD-CSI computation. By extending $\Delta t$, the contribution of the dynamic component in TD-CSI becomes more pronounced, thereby mitigating the impact of noise and improving the robustness of SSNR measurement. As shown in Figure~\ref{fig3_6}, increasing $\Delta t$ results in more stable phase differences in TD-CSI, highlighting the importance of selecting an appropriate $\Delta t$. If $\Delta t$ is too short, all subcarriers may be dominated by noise, leading to distorted SSNR estimation. However, in real-world scenarios, human movement speeds vary dynamically, making it impractical to use a fixed $\Delta t$ that fits all cases. To address this, we adopt a multi-interval approach. Specifically, we construct multiple TD-CSI signals using different time intervals. For each subcarrier, we compute the variance of phase differences across these intervals and take the average variance as the final SSNR value. This strategy effectively enhances SSNR reliability, mitigates the impact of noise, and generalizes to diverse movement speeds, ensuring robust SSNR quantification and angle estimation performance across various real-world scenarios.

\section{Implementation}

\begin{figure}[!b]
    \centering
        \subfloat[The fabricated FSA]{
        \includegraphics[height=0.8in]{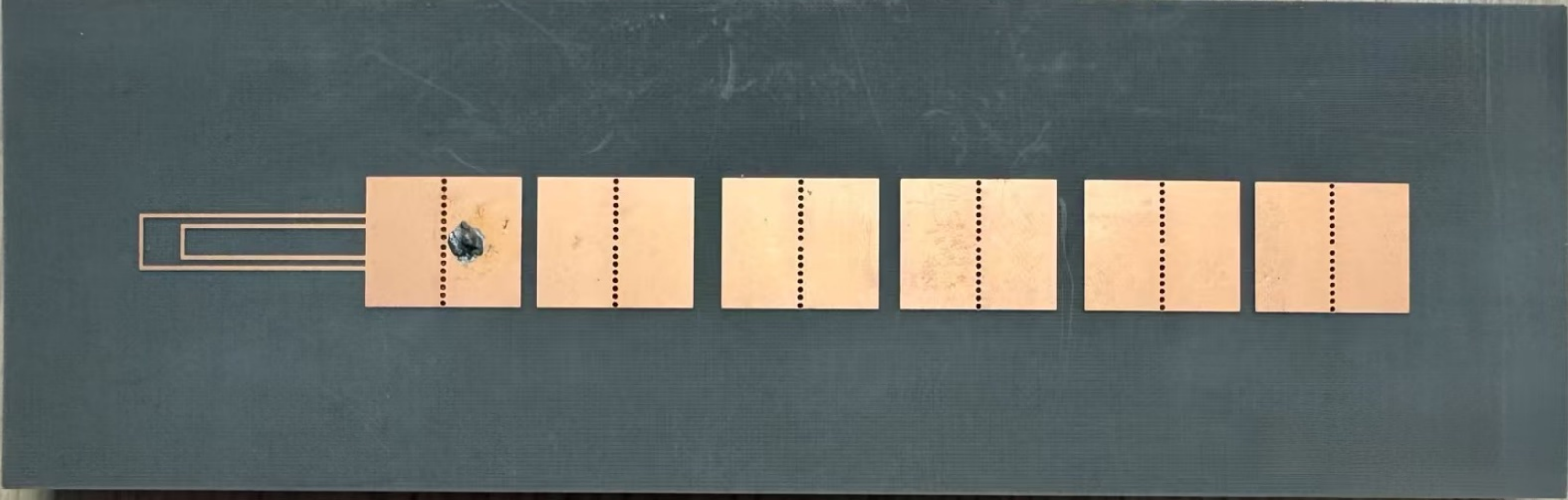}
        \label{fig4_1a}
        \hspace{0in}
    }\\
        \subfloat[Parameters of antenna dimension]{
        \includegraphics[height=0.9in]{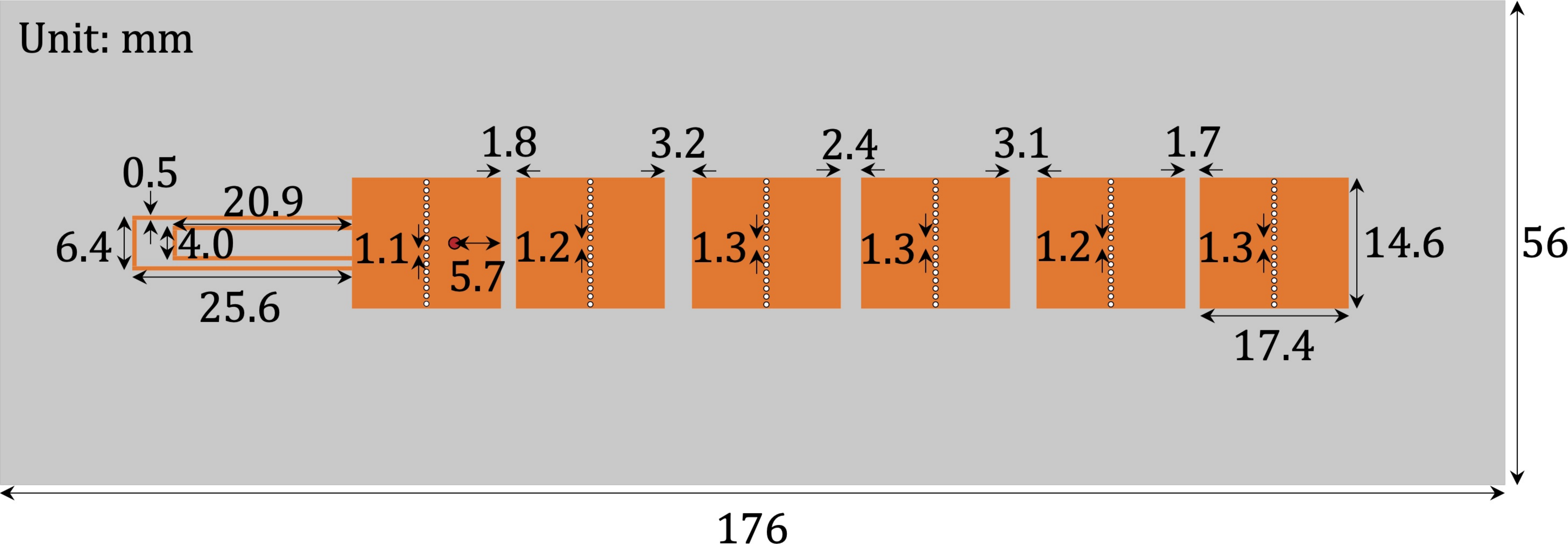}
        \label{fig4_1b}
        \hspace{0in}
    }
    \vspace{-0em}
    \caption{Hardware Implementation.}
    \label{fig4_1}
\end{figure}

\subsection{Hardware Implementation}\label{sec41}
The prototype of the proposed FSA is shown in Figure~\ref{fig4_1a}. Electromagnetic simulation and antenna optimization are performed with the commercial software of Altair Feko 2019~\cite{feko} and CST Microwave Studio~\cite{cst}. The antenna measures 17.4 cm $\times$ 5.5 cm and is fabricated on a double-sided PCB using a 0.787 mm thick Wangling F4BTMS220 substrate~\cite{wanglingF4BTMS}, with a dielectric constant of 2.2 and loss tangent of 0.0009. This PTFE-based ceramic-filled material offers excellent frequency stability, low loss, and low anisotropy, making it well-suited for antenna design. The metalized via holes have a radius of 0.5 mm, with a 0.5 mm edge-to-edge spacing between adjacent vias. Detailed dimensions of the antenna structure are provided in Figure~\ref{fig4_1b}. A feed port is located on the back for connection to a Wi-Fi transmitter. One of the key advantages of the proposed FSA is its low cost, as it utilizes standard PCB fabrication materials and does not require expensive RF control circuits or multiple RF transceiver chains. For reception, the system uses commodity omnidirectional Wi-Fi antennas. For signal transmission and reception, we employ two USRP B210s, with both transceivers connected to a laptop for real-time signal transmission and CSI extraction. Note that the resonator-based design can be easily adapted to other frequency bands by adjusting the physical dimensions and material properties of the antenna.

\begin{figure}[!t]
    \centering
        \subfloat[Beam directions across frequencies]{
        \includegraphics[height=1.2in]{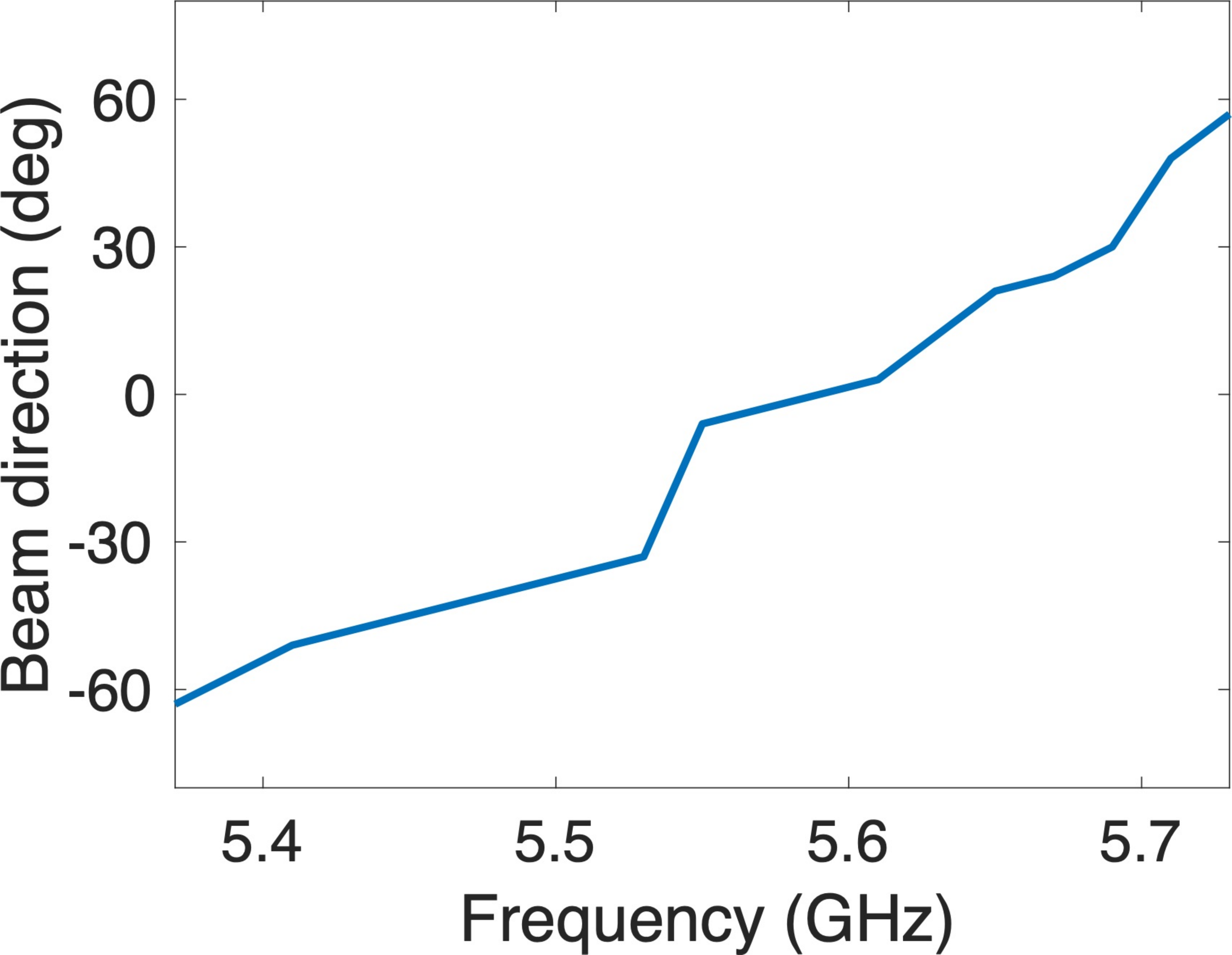}
        \label{fig4_3a}
        \hspace{0in}
    }
    \subfloat[Gains of five directions across frequencies]{
        \includegraphics[height=1.2in]{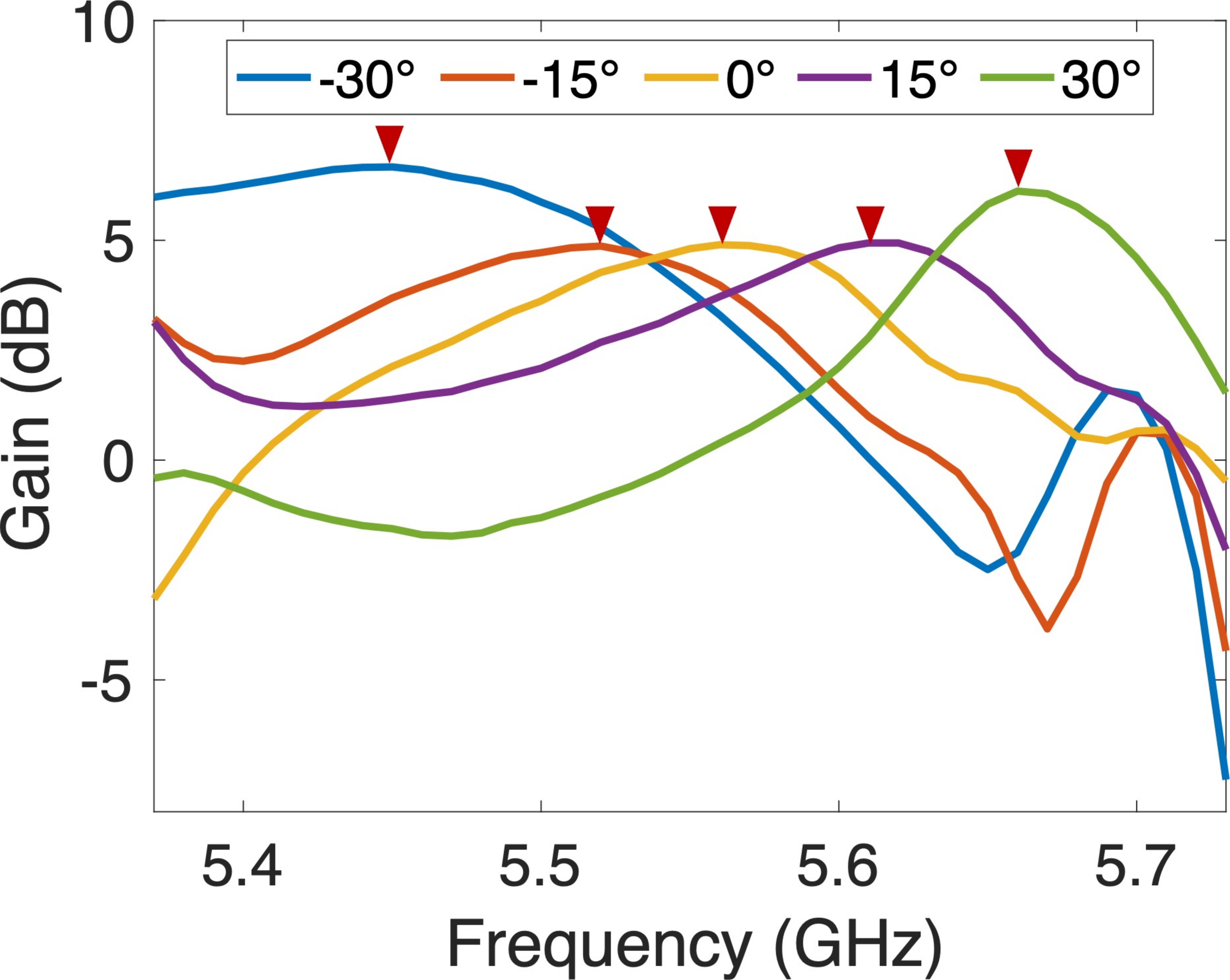}
        \label{fig4_3b}
        \hspace{0in}
    }
    \vspace{-0em}
    \caption{Measurement results in an anechoic chamber.}
    \label{fig4_3}
\end{figure}

To validate the frequency-scanning characteristics of the fabricated FSA, we conduct radiation measurements in an anechoic chamber. Figure~\ref{fig4_3a} shows the measured beam direction for each frequency. The result indicates that the antenna achieves a FoV of approximately 120$\degree$ over the full 360 MHz bandwidth from 5.37 GHz to 5.73 GHz. However, due to Wi-Fi protocol constraints, the maximum usable bandwidth within this frequency band is limited to 160 MHz (i.e., Channel 114, 5.49 – 5.65 GHz), within which the effective FoV is about 60$\degree$ (i.e., from -21$\degree$ to 39$\degree$). Within this FoV, we further measure the gain of each frequency at five representative directions (i.e., from -30$\degree$ to 30$\degree$). As shown in Figure~\ref{fig4_3b}, each angle corresponds to a distinct frequency at which the antenna gain is maximized, demonstrating the frequency–direction mapping property of the FSA. These measurements serve as a useful guide for the evaluation experiments. 

\vspace{-0.0em}
\subsection{Software Implementation}
\begin{figure}[!t]
\vspace{-0.0em}
	\includegraphics[height=1.2in]{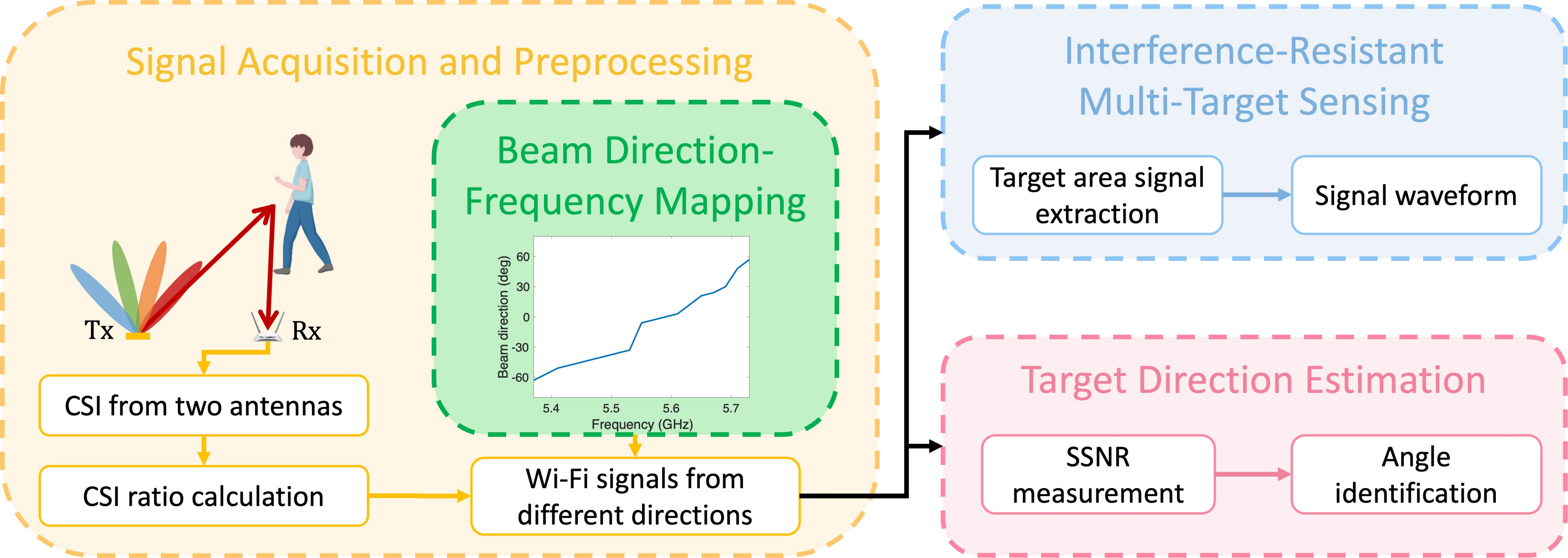}
	\centering{}
	\caption{\systemname{} overview.}
	\label{fig4_2}
    \vspace{-0.0em}
\end{figure}

As illustrated in Figure~\ref{fig4_2}, our system comprises three key modules. The first module extracts and calibrates Wi-Fi CSI from the received packets, enabling phase-corrected data for subsequent sensing tasks. The processed CSI can then be used either for interference-resistant multi-target sensing or for estimating the direction of a target within the FoV.

\vspace{-0.0em}

\subsubsection{Signal Acquisition and Preprocessing}
For Wi-Fi signal transmission and CSI extraction, we employ PicoScenes~\cite{jiang2021eliminating}, a widely used Wi-Fi sensing platform that supports packet injection and CSI collection. PicoScenes operates on Ubuntu 20.04, while CSI processing and higher-level signal analysis are performed in MATLAB 2024b. In our setup, CSI is simultaneously collected from two receiving antennas at a packet rate of 200 Hz, providing sufficient temporal resolution for motion and respiration sensing. Because Wi-Fi transceivers are not synchronized, the raw CSI phase is affected by random offsets. To compensate for this, we adopt the CSI ratio method~\cite{zeng2019farsense}, which divides the CSI from one receiving antenna by that from the other. Since both antennas share identical hardware-induced phase offsets, their ratio effectively removes the offset, resulting in clean, phase-stable CSI. After calibration, we utilize the beam direction–frequency mapping obtained in Section~\ref{sec41}. This mapping allows us to associate each subcarrier with a specific direction within the FoV, enabling subcarrier-level spatial separation of signals corresponding to different angles. The resulting set of calibrated subcarrier signals forms the input to the next two modules.

\vspace{-0.0em}

\subsubsection{Interference-Resistant Multi-Target Sensing}
In many real-world environments, signals reflected from multiple objects coexist, causing interference and reducing sensing accuracy. To mitigate this, we exploit the frequency–direction mapping of FSAs to isolate signals originating from different angular regions. For each subcarrier, the CSI ratio signal represents reflections mainly from the corresponding beam direction. By selecting the subcarriers whose beam directions match the desired sensing area, the system can selectively extract motion-related signals from that direction while suppressing reflections from others. This mechanism naturally enables interference-resistant and multi-target sensing without requiring multi-antenna beamforming.

\begin{figure*}[!b]
    \centering
        \subfloat[Experiment setup]{
        \includegraphics[height=1.25in]{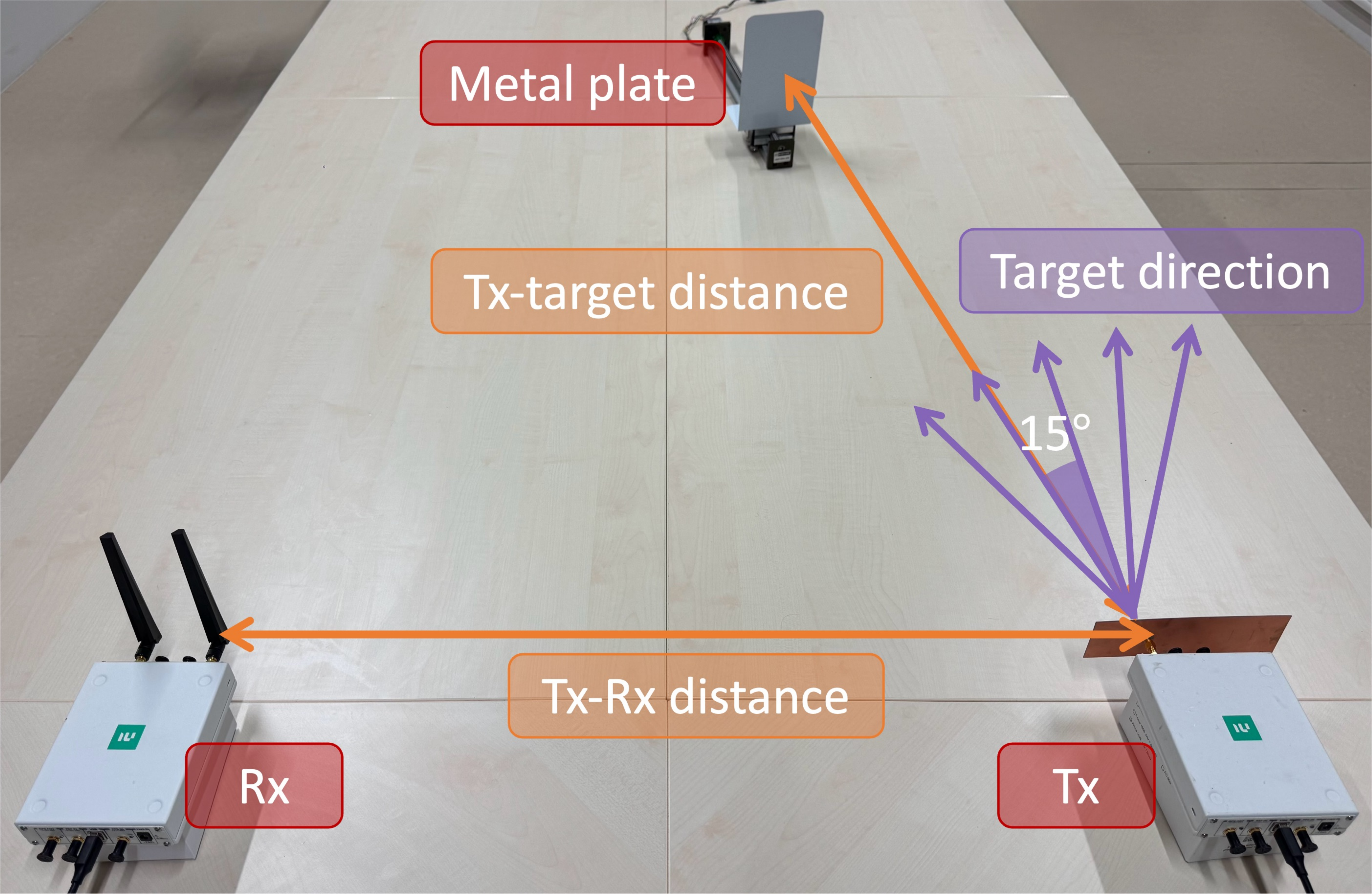}
        \label{fig5_1a}
        \hspace{0in}
    }
        \subfloat[The impact of different angle between FSA and target]{
        \includegraphics[height=1.25in]{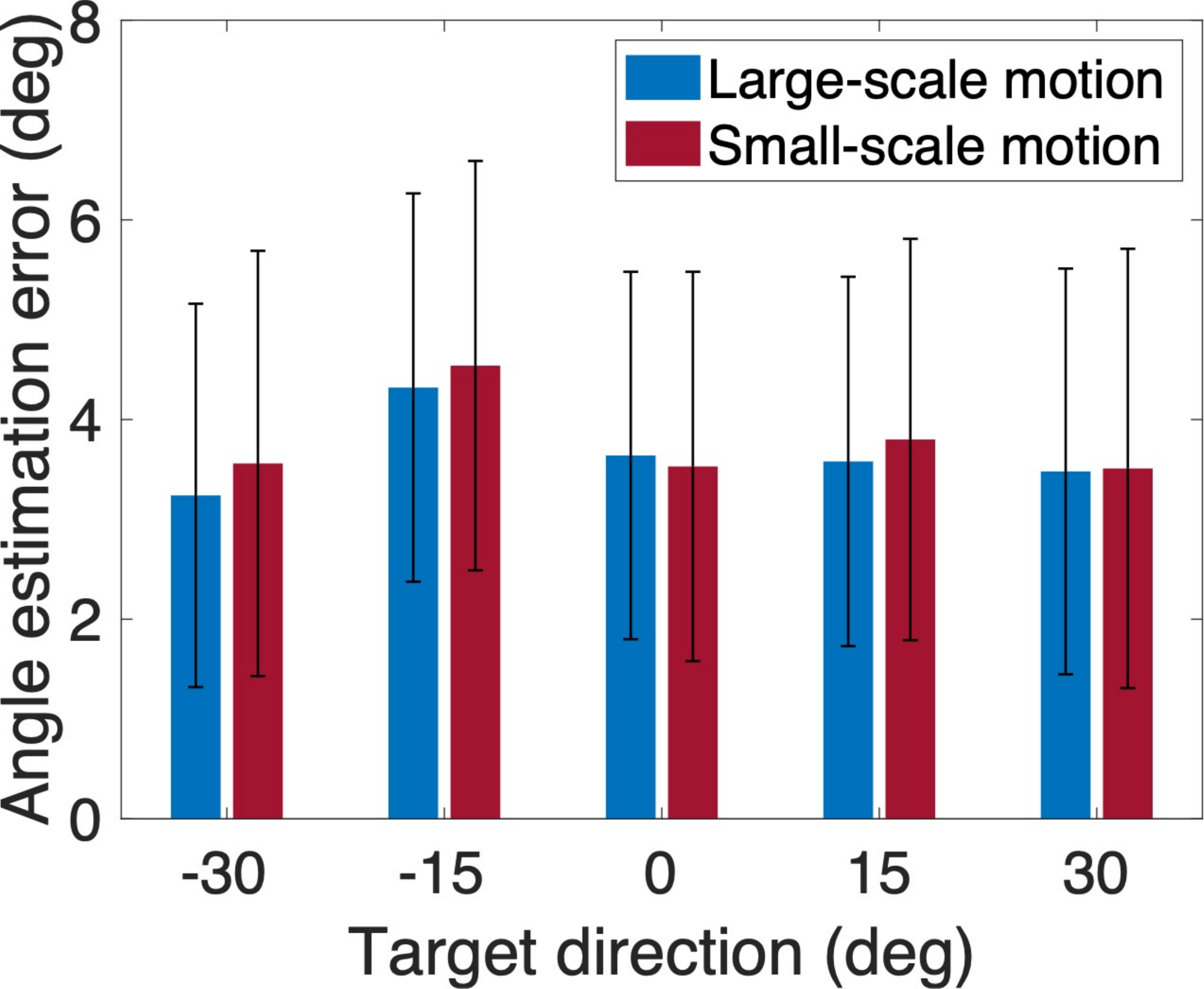}
        \label{fig5_1c}
        \hspace{0in}
    }
        \subfloat[The impact of Tx-target distance]{
        \includegraphics[height=1.25in]{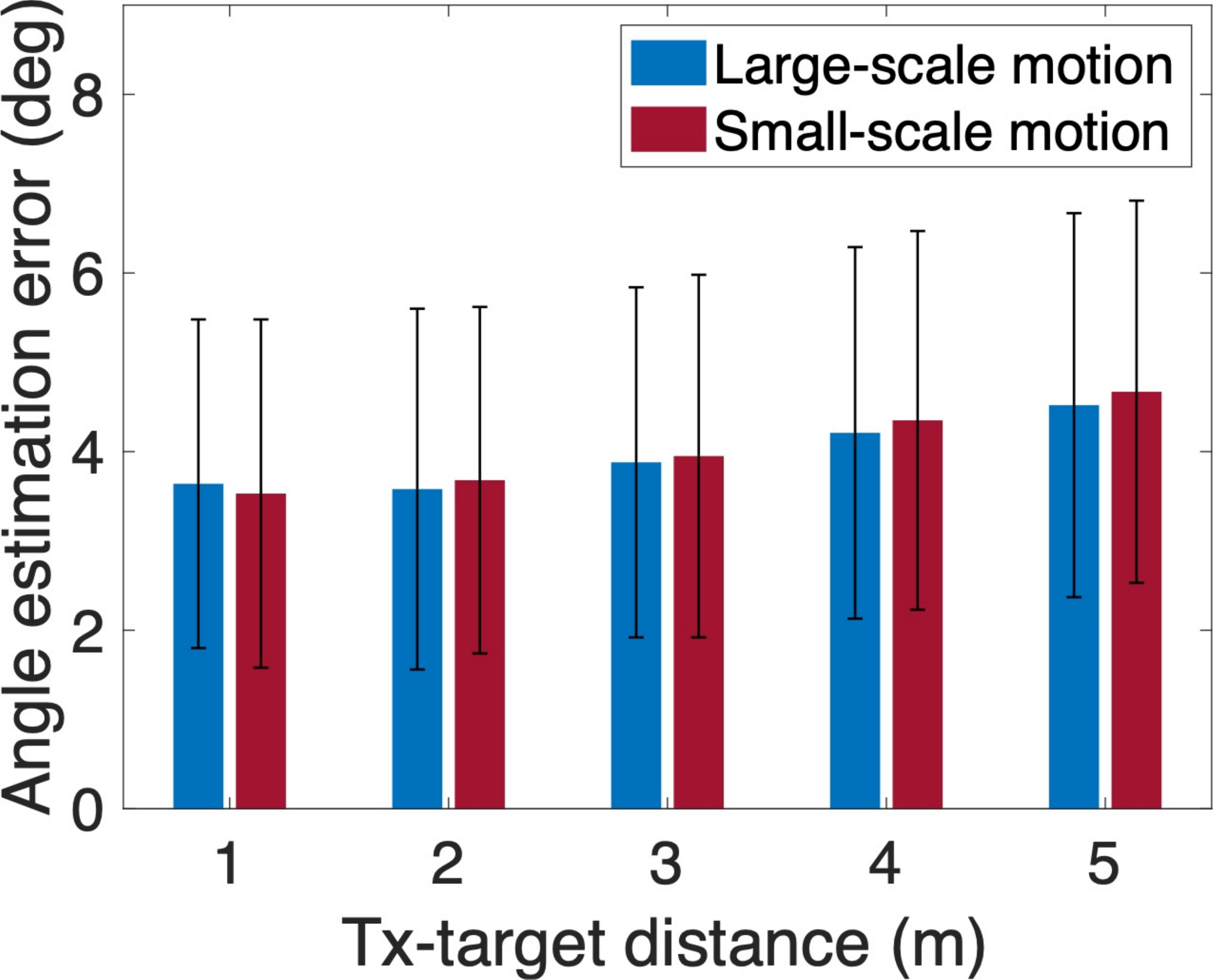}
        \label{fig5_1d}
        \hspace{0in}
    }
    \subfloat[The impact of transceiver distance]{
        \includegraphics[height=1.25in]{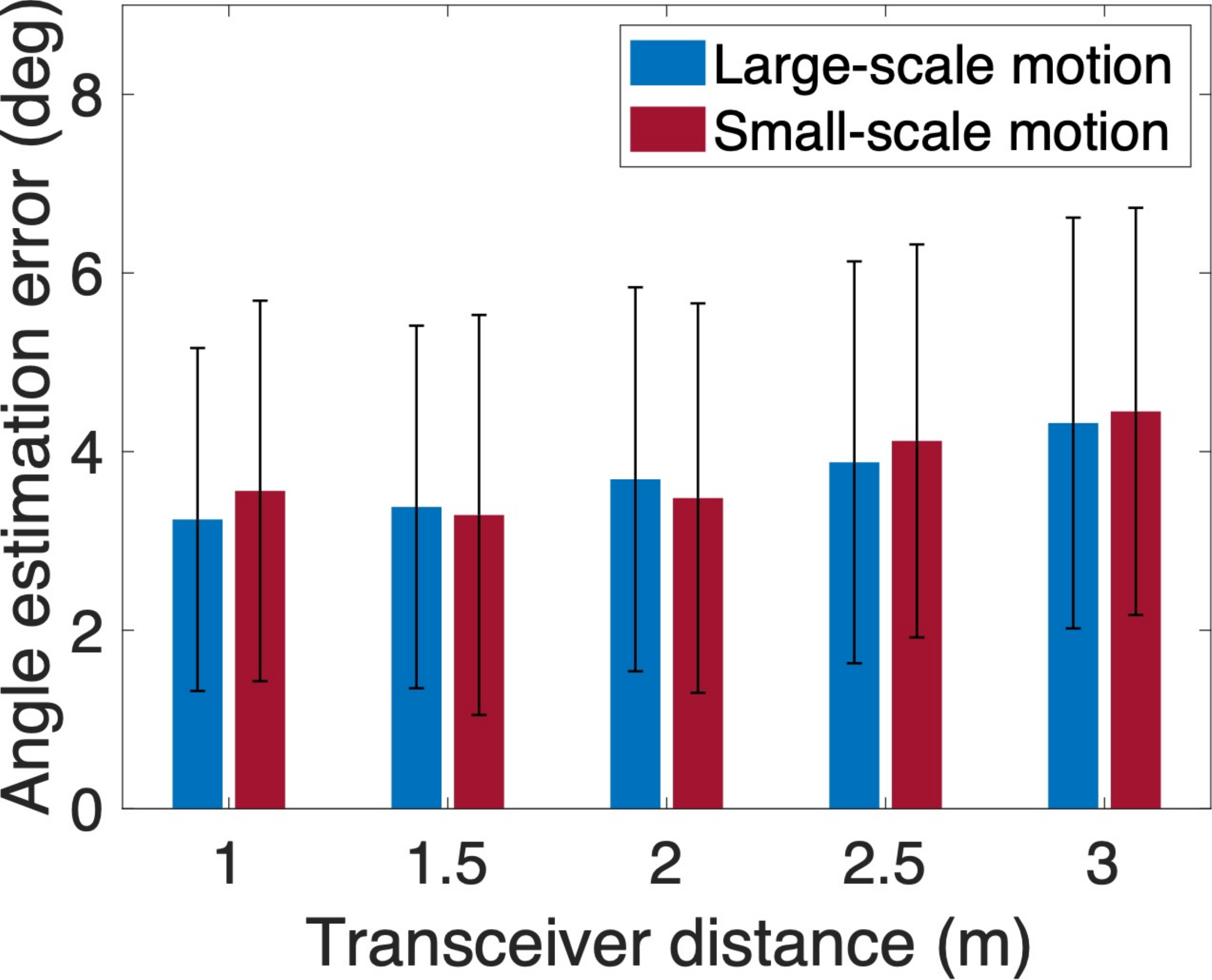}
        \label{fig5_1e}
        \hspace{0in}
    }
    \vspace{-0em}
    \caption{Benchmark experiments.}
    \label{fig5_1n}
\end{figure*}

\subsubsection{Target Direction Estimation}
For direction estimation, we first construct multiple sub-sampled signals across a range of time intervals. The time interval $\Delta t$ is varied between 0.005 s and 0.1 s to capture dynamic components over multiple motion speeds. Given the CSI sampling rate of 200 Hz, a total of 20 possible intervals can be used to generate TD‑CSI signals. We then compute TD‑CSI for all intervals to eliminate the static component, as described in Section~\ref{sec32}. For each TD‑CSI signal, we quantify the stability of phase variations by calculating the variance of consecutive phase differences $\Delta(\angle\Delta H(f,t))$. This variance serves as an indicator of phase smoothness and SSNR. For each subcarrier, we average the variances obtained across all intervals to determine the final SSNR value. The subcarrier with the highest SSNR corresponds to the beam direction most strongly affected by the target’s movement. Since each subcarrier is uniquely associated with a direction through the established frequency–direction mapping, the corresponding angle of this subcarrier is identified as the target’s direction.

\vspace{-0.0em}
\section{Evaluation}

In this section, we evaluate the capability of direction-aware sensing performance of \systemname{} through benchmark experiments and two real-world case studies. Benchmark experiments are conducted to validate the effectiveness of \systemname{} under various conditions for direction estimation, including the impact of device placement and the position relationship between the antenna and the target. In case studies, we showcase that \systemname{} can be deployed to achieve direction-aware sensing in several real-world scenarios for different purposes, including walking area detection and multi-target interference-resistant respiration monitoring.

\subsection{Benchmark Experiments}

\subsubsection{Experiment Setting}
The goal of the benchmark experiments is to validate the direction estimation capability of the proposed FSA prototype. Specifically, we aim to demonstrate that the frequency-dependent angular dispersion effect, previously verified through anechoic chamber measurements in Section~\ref{sec41}, can be effectively leveraged for direction-aware sensing. As illustrated in Figure~\ref{fig5_1a}, we conduct the experiments in a large meeting room using a metallic plate as the reflective target. A Raspberry Pi-controlled sliding track is employed to precisely control the target’s movement, while the Wi-Fi transmitter and receiver are placed 1 meter apart by default. The FSA is oriented toward the region of interest, and the target is sequentially positioned at five angular locations ranging from –30$\degree$ to +30$\degree$ in 15$\degree$ increments, covering the antenna’s 60$\degree$ FoV at Wi-Fi channel 114. Since the USRP B210 used in our setup supports only a limited instantaneous bandwidth, each test is conducted multiple times, with the operating frequency adjusted across adjacent sub-bands to span the entire 160 MHz band. All experiments are repeated ten times to ensure statistical reliability.
To comprehensively evaluate performance under different motion scales, we consider two representative motion patterns:
\begin{itemize}[leftmargin=*]
\item \textbf{Large-scale motion:} The target moves 1 m at a speed of 10 cm/s, representing scenarios such as human tracking or gesture recognition.
\item \textbf{Small-scale motion:} The target oscillates periodically with 1 cm displacements at a speed of 1 cm/s, simulating subtle motions such as respiration.
\end{itemize}

\begin{figure*}[!b]
    \centering
        \subfloat[S1: A corner in a smart building]{
        \includegraphics[height=1.1in]{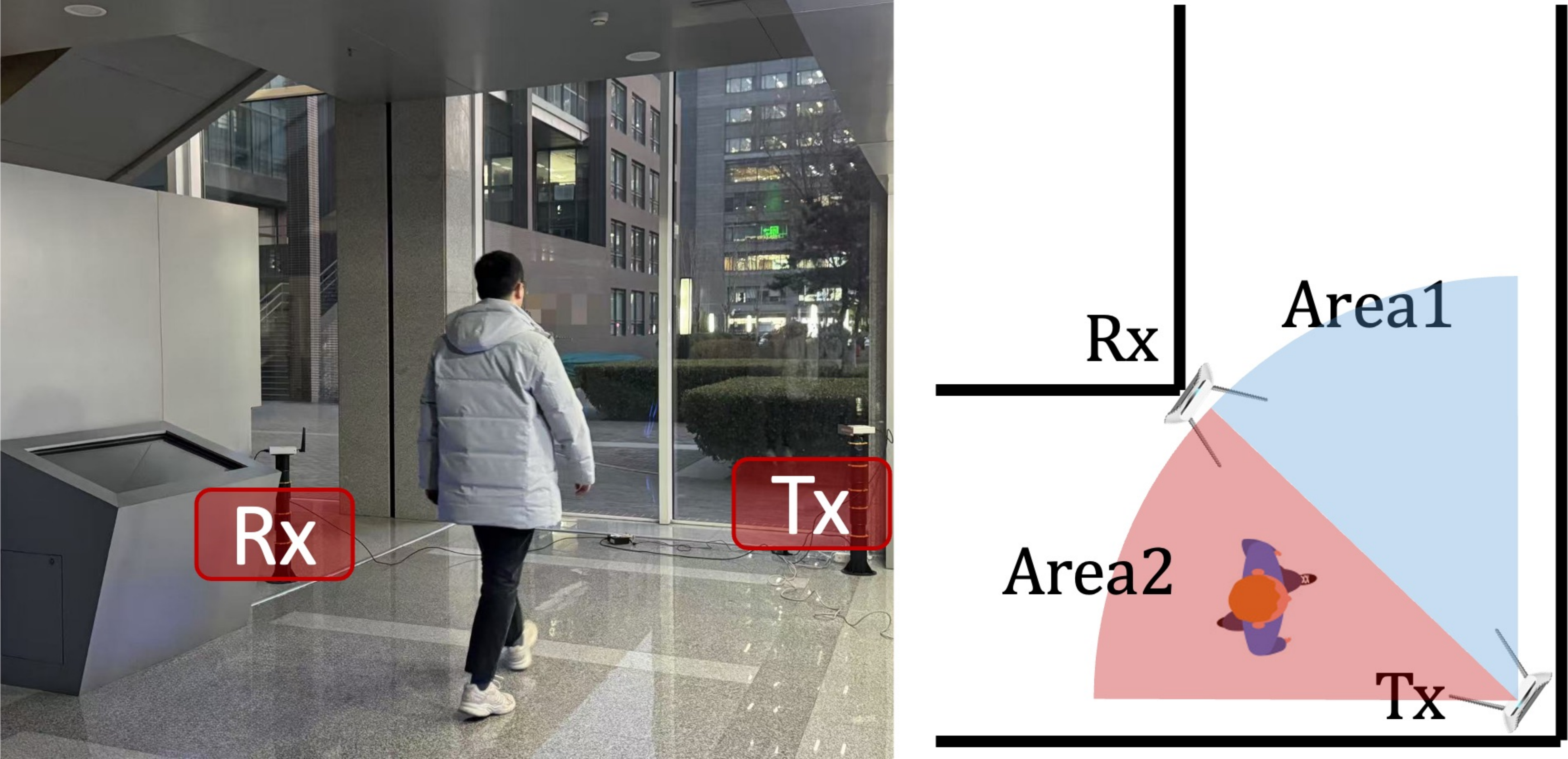}
        \label{fig5_2a}
        \hspace{0in}
    }
        \subfloat[S2: An entrance in a smart building]{
        \includegraphics[height=1.15in]{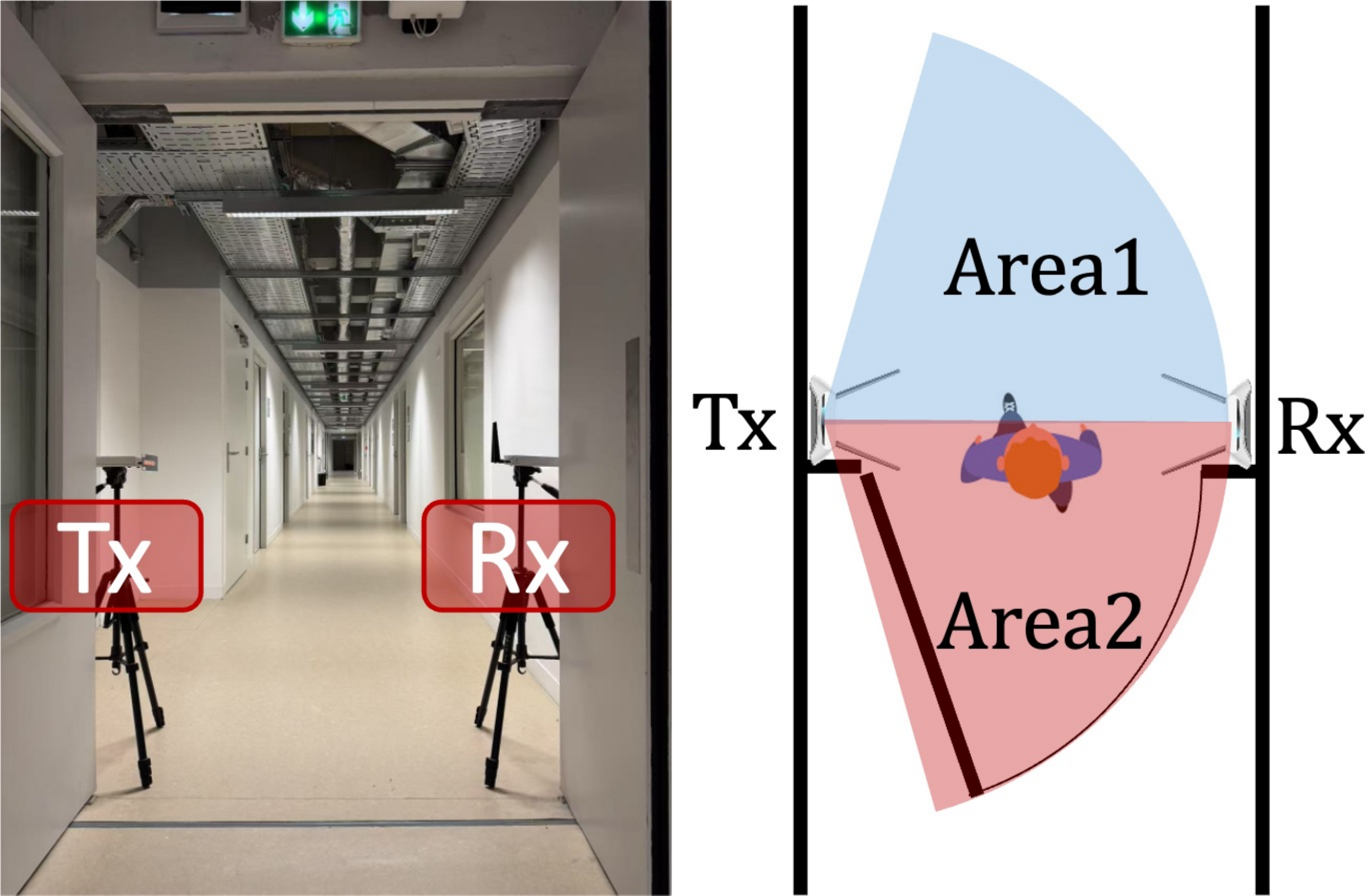}
        \label{fig5_2b}
        \hspace{0in}
    }
        \subfloat[S3: A living room in a smart home]{
        \includegraphics[height=1.25in]{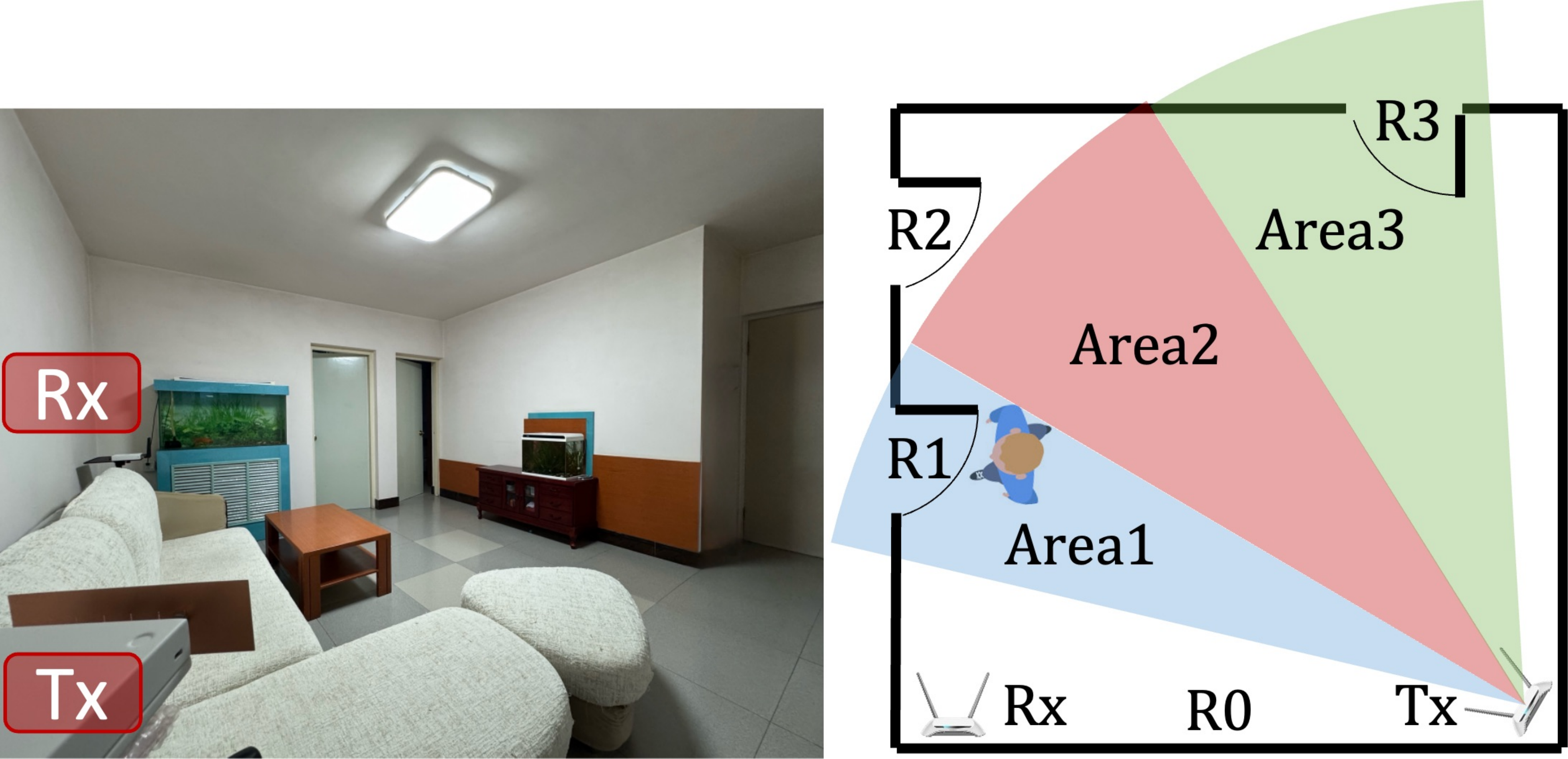}
        \label{fig5_2c}
        \hspace{0in}
    }
    \vspace{-0.0em}
    \caption{Scenarios of direction-aware walking area detection.}
    \label{fig5_2}
    \vspace{-0em}
\end{figure*}

\subsubsection{Overall Performance}
Figure~\ref{fig5_1c} shows the angle estimation accuracy across five target directions. The average estimation error remains below 4.4$\degree$ for large-scale motion and 4.6$\degree$ for small-scale motion. In addition, the accuracy exhibits minimal variation across different directions, indicating that the proposed system achieves stable and reliable direction estimation across the entire angular range and for both motion types.

\subsubsection{Impact of Target Distance}
In this experiment, we vary the target distance from the FSA from 1 m to 5 m in increments of 1 m. The target remains fixed at 0$\degree$ within the FoV. As shown in Figure~\ref{fig5_1d}, the estimation error increases slightly as the distance grows but remains below 4.6$\degree$ for large-scale motion and 4.7$\degree$ for small-scale motion. These results demonstrate that the proposed method maintains robust performance over practical sensing distances.

\subsubsection{Impact of Transceiver Separation}
We further investigate the effect of the separation distance between the Wi-Fi transmitter and receiver, which is varied from 1 m to 3 m in increments of 0.5 m. The target is kept at 0$\degree$ within the antenna FoV. As shown in Figure~\ref{fig5_1e}, the estimation error remains consistently below 4.4$\degree$ and 4.5$\degree$ for the two motion types, respectively. This experiment confirms that the proposed system maintains stable accuracy under different transceiver placements, demonstrating its practicality for real-world deployment.

\begin{figure*}[!t]
    \centering
        \subfloat[Four walking trajectories]{
        \includegraphics[height=1.35in]{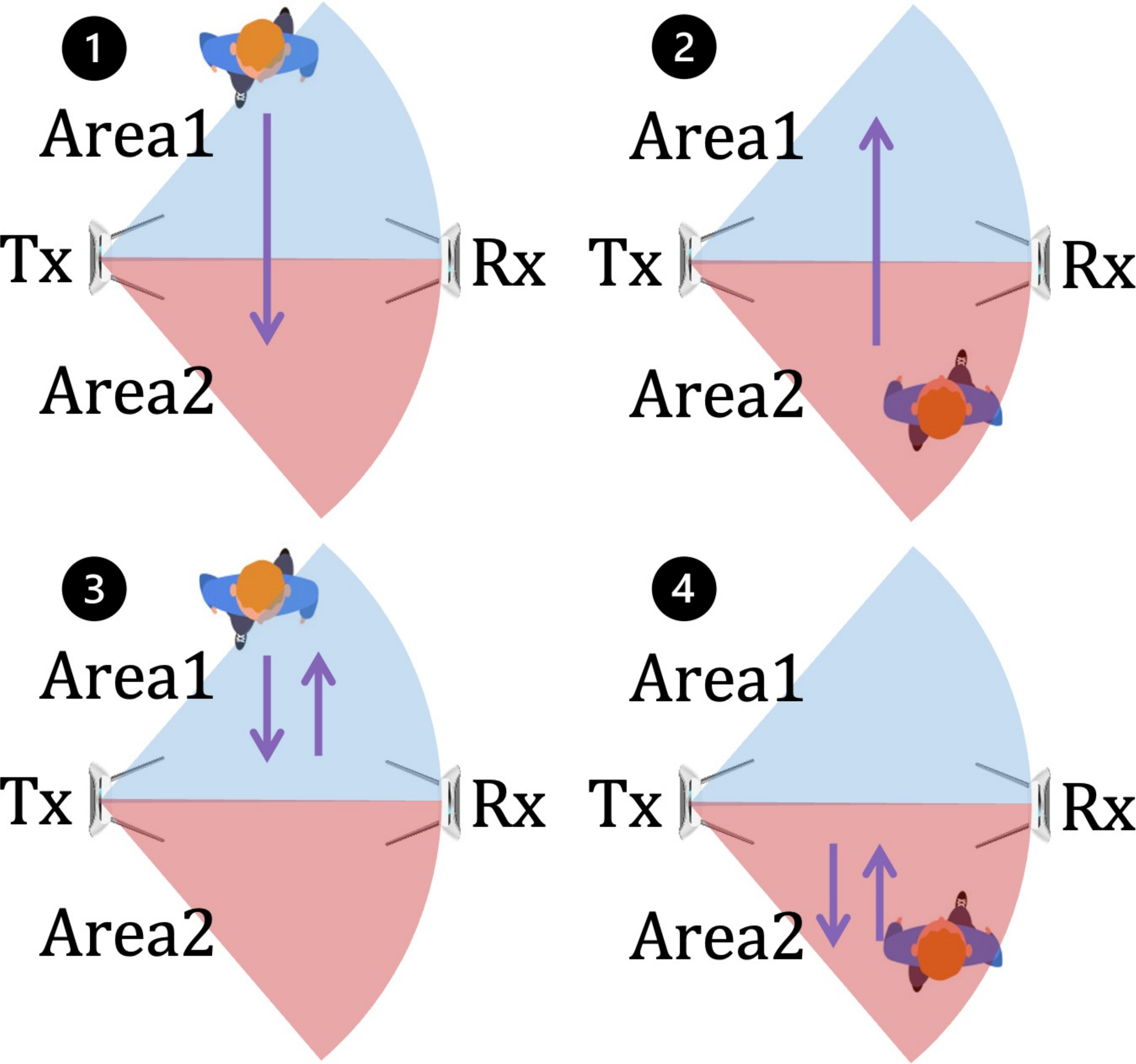}
        \label{fig5_3a}
        \hspace{0in}
    }
        \subfloat[Confusion matrix]{
        \includegraphics[height=1.35in]{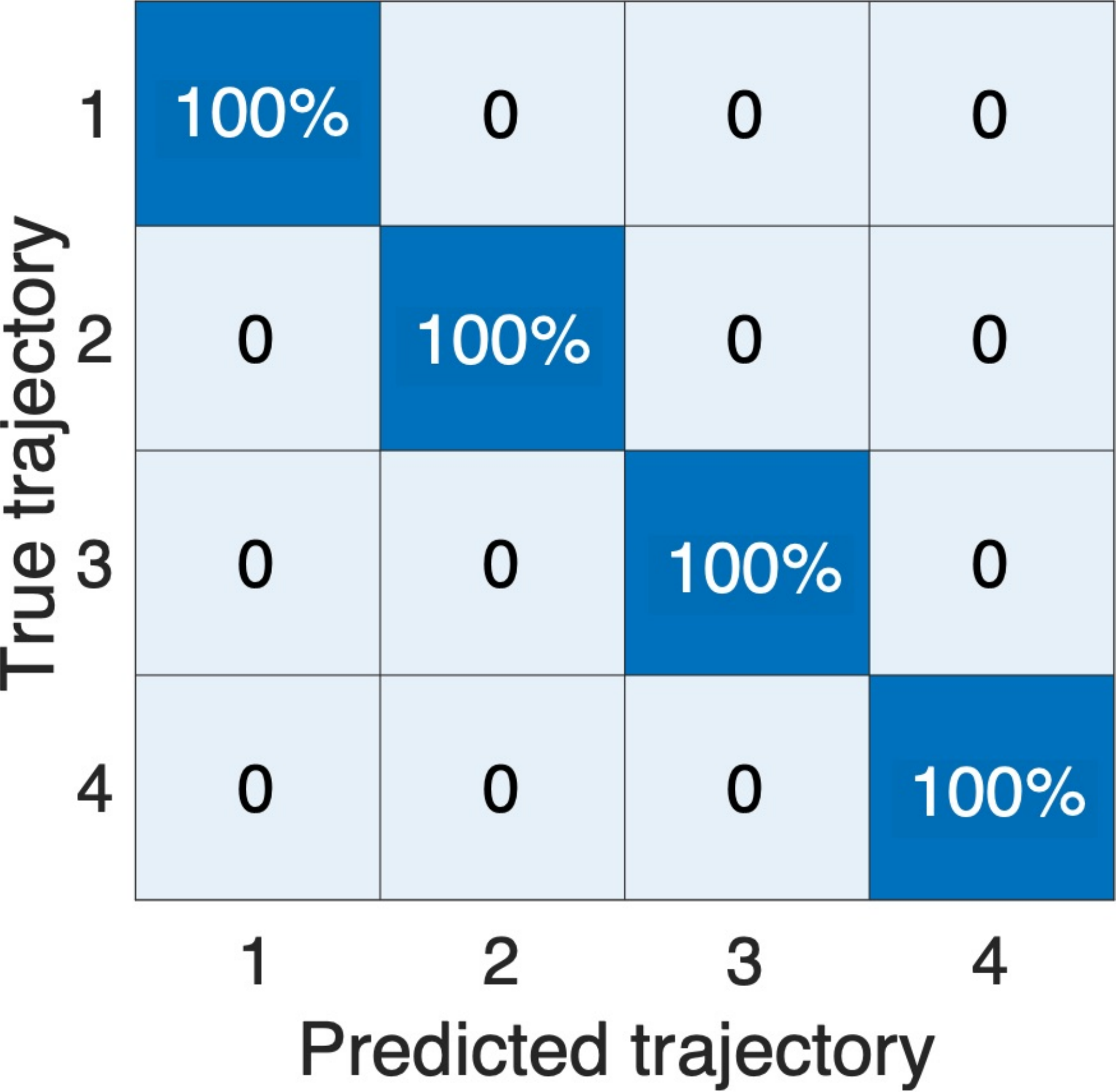}
        \label{fig5_3b}
    }
        \hspace{0.2in}
        \subfloat[Device placement in S3]{
        \includegraphics[height=1.35 in]{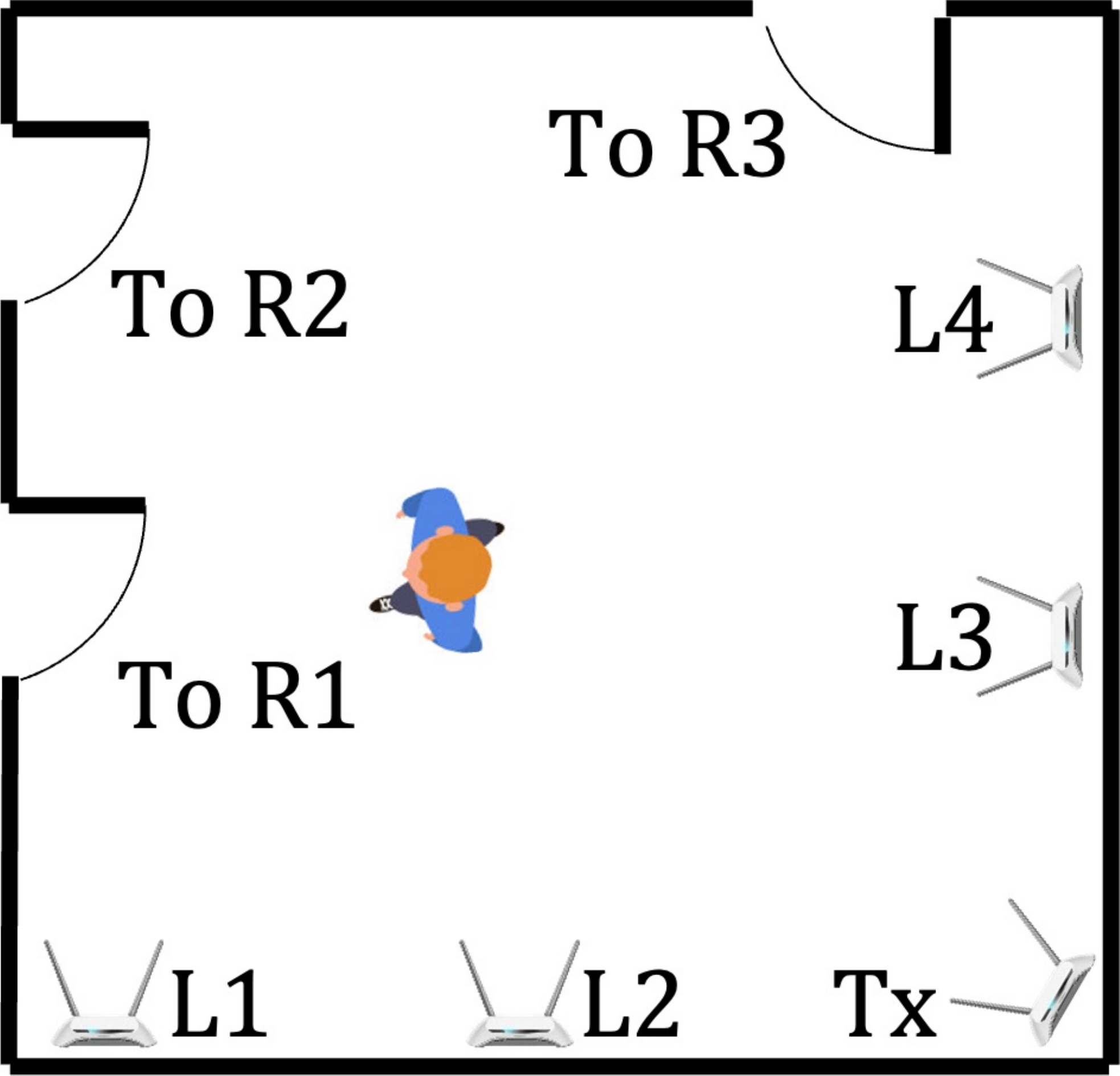}
        \label{fig5_3c}
        \hspace{0in}
    }
    \subfloat[Result of room detection detection]{
        \includegraphics[height=1.35in]{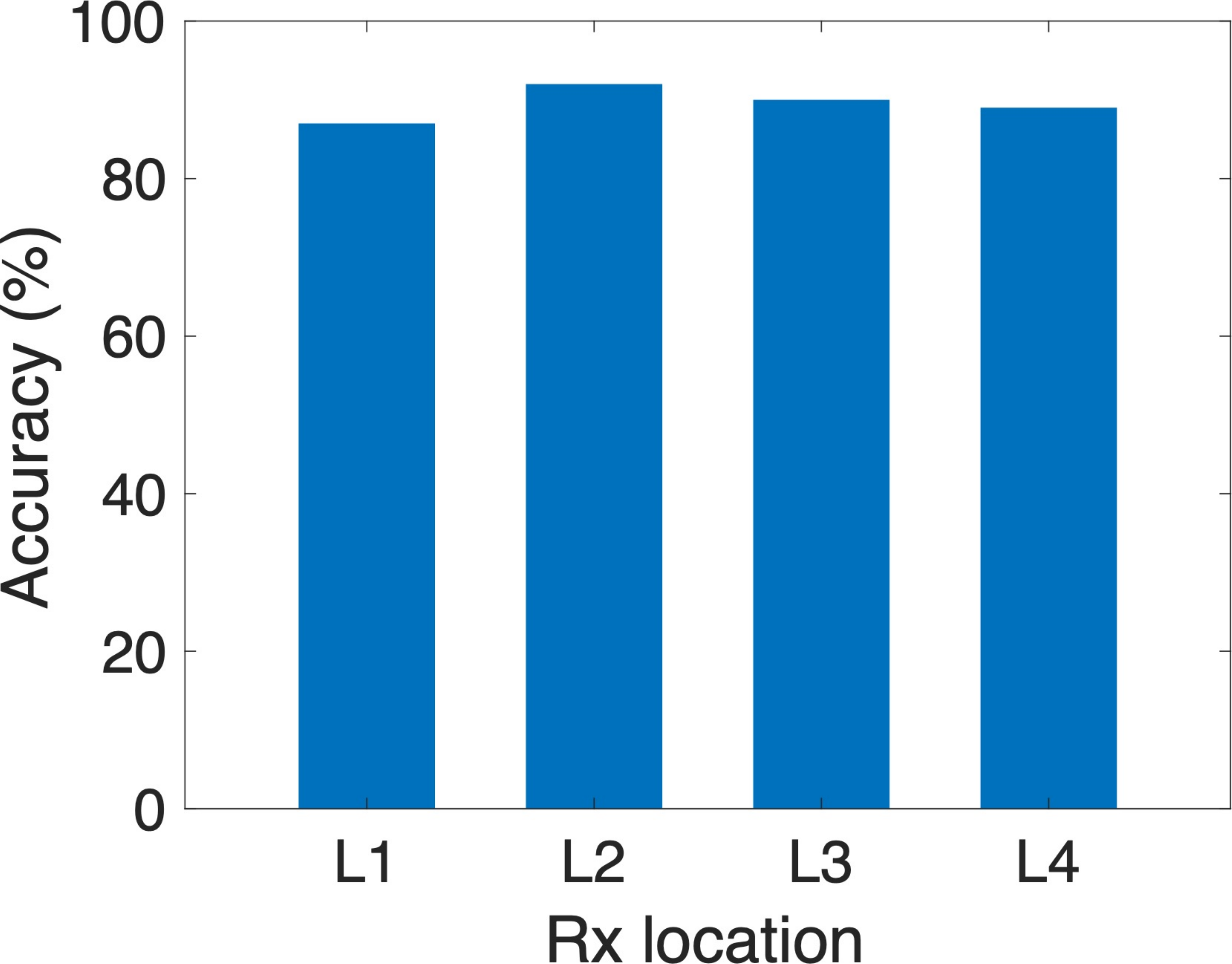}
        \label{fig5_3d}
        \hspace{0in}
    }
    \vspace{-0em}
    \caption{Evaluation on direction-aware walking area detection based on angle estimation.}
    \vspace{-0em}
    \label{fig5_3}
\end{figure*}

\vspace{-0.0em}
\subsection{Case Study 1: Direction-Aware Walking Area Detection with Angle Estimation}
Understanding which area a person is walking through is essential for enabling context-aware smart environments. Wi-Fi sensing has shown strong potential for this task, particularly through walking trajectory reconstruction, relying on antenna arrays or multiple transceiver pairs~\cite{li2017indotrack,niu2022rethinking,li2024wifi}. In this section, we propose an alternative approach that uses only a single transceiver pair and one FSA transmitting antenna. Rather than reconstructing the full trajectory, we divide the environment into discrete regions based on their angular positions relative to the FSA. By estimating the target’s direction during walking, the system can determine which region the person is passing through, providing location-aware contextual information while maintaining a lightweight and cost-efficient deployment suitable for a wide range of applications.

\begin{figure*}[!b]
    \centering
        \subfloat[Multi-target respiration monitoring]{
        \includegraphics[height=1.4in]{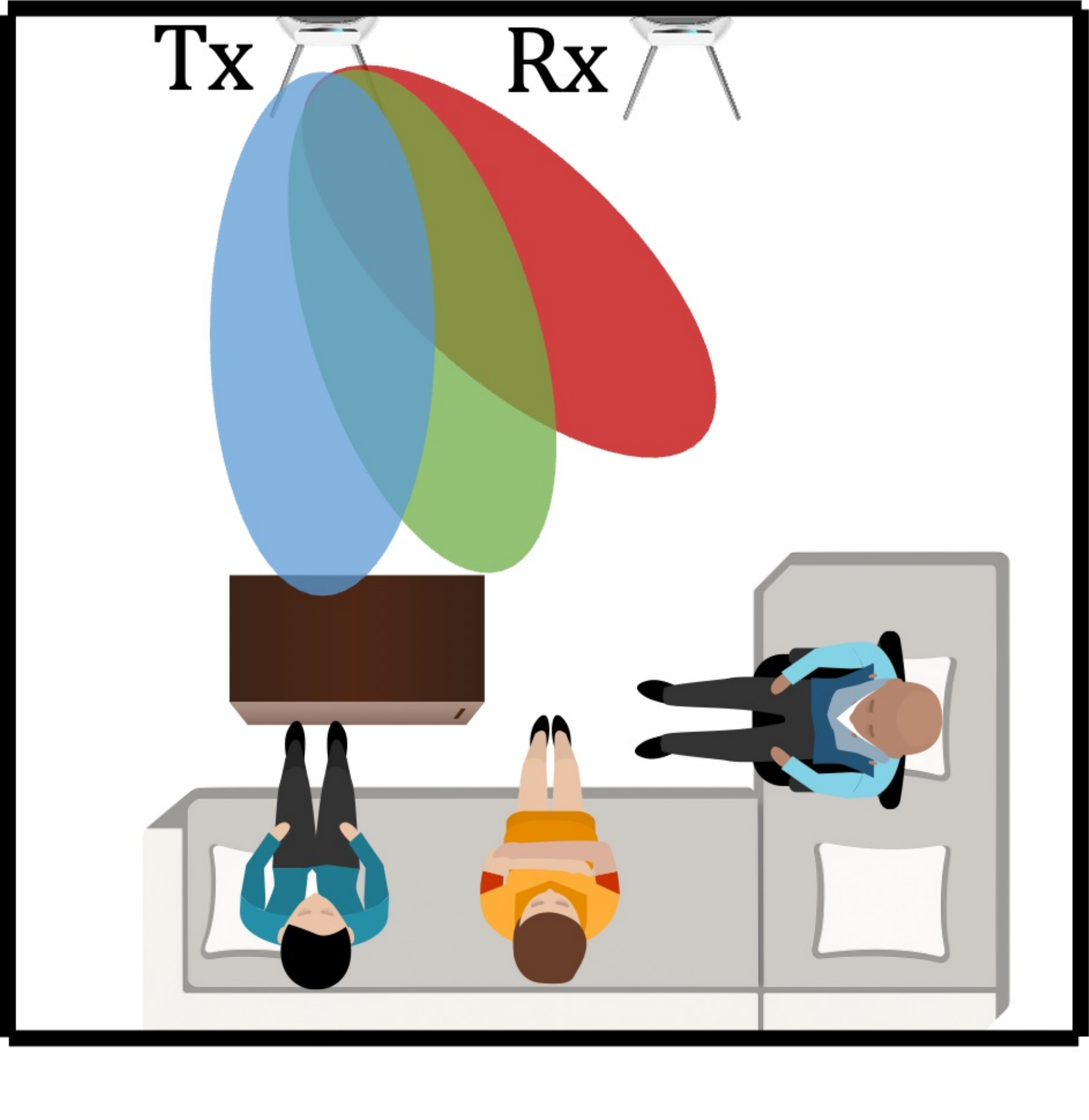}
        \label{fig5_5a}
        \hspace{0in}
    }
        \subfloat[Respiration rate estimation error]{
        \includegraphics[height=1.4in]{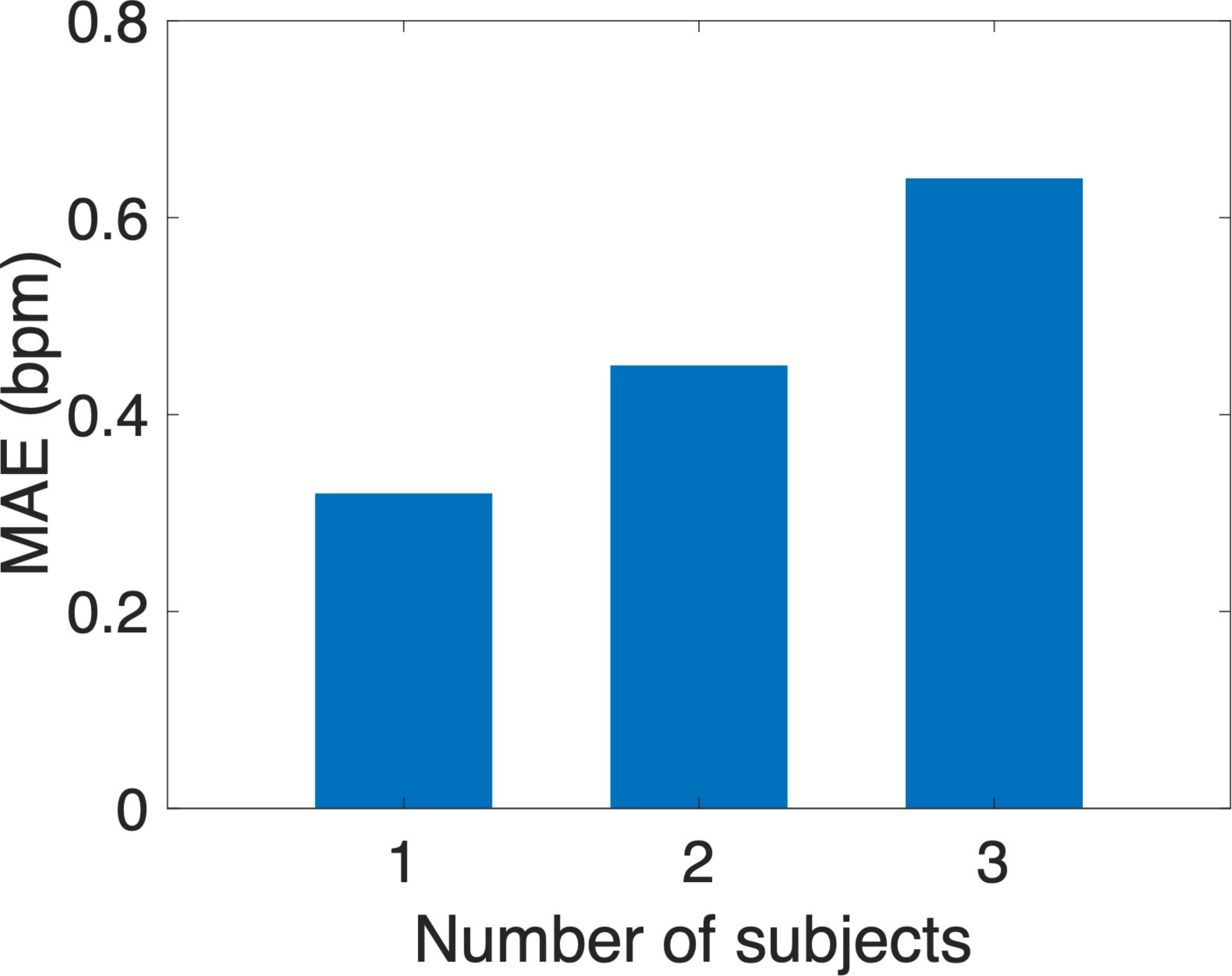}
        \label{fig5_5b}
    }
        \hspace{0.0in}
        \subfloat[Interference from a walking person]{
        \includegraphics[height=1.4in]{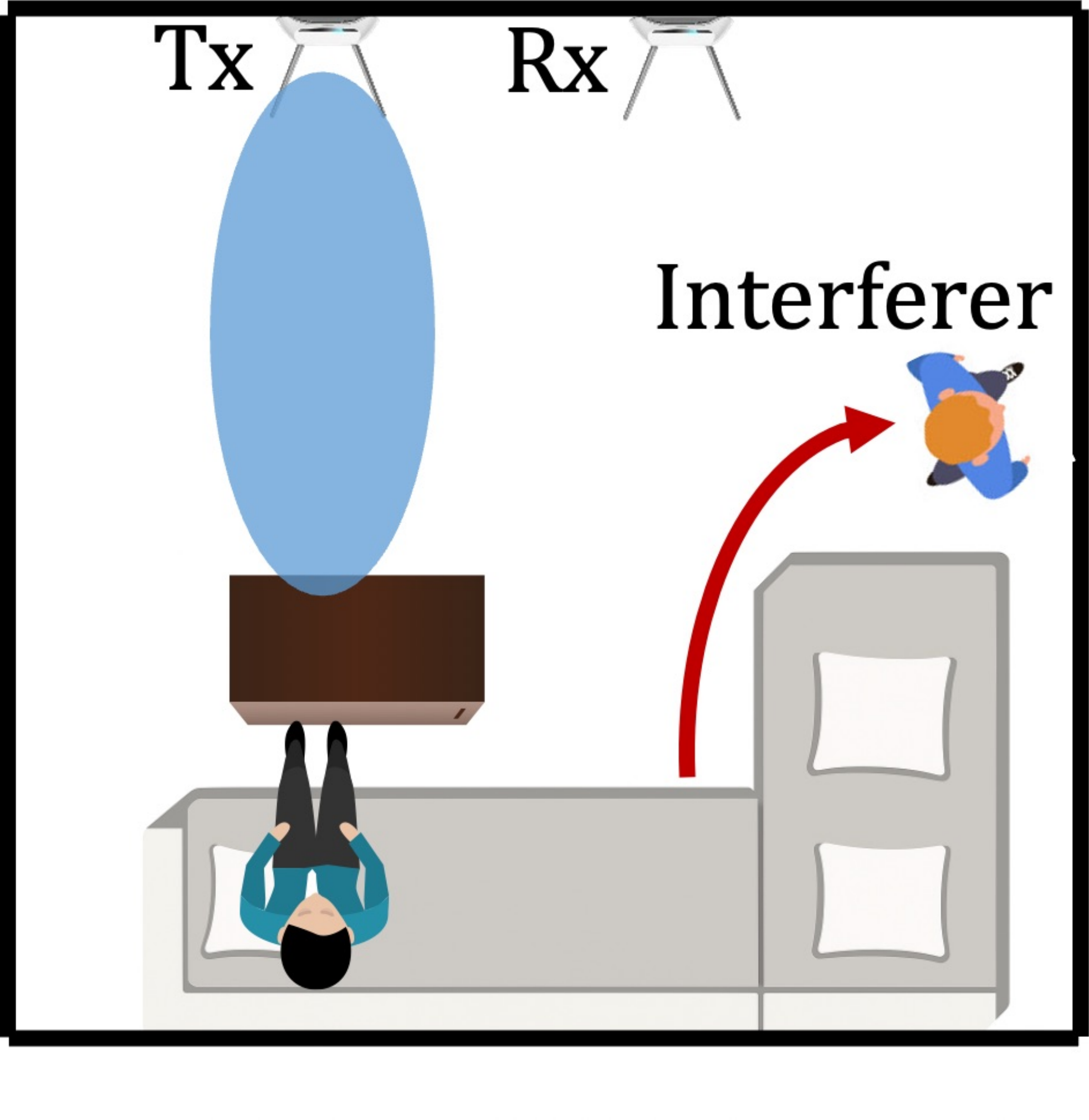}
        \label{fig5_6a}
        \hspace{0in}
    }
    \subfloat[Comparison between two antennas]{
        \includegraphics[height=1.4in]{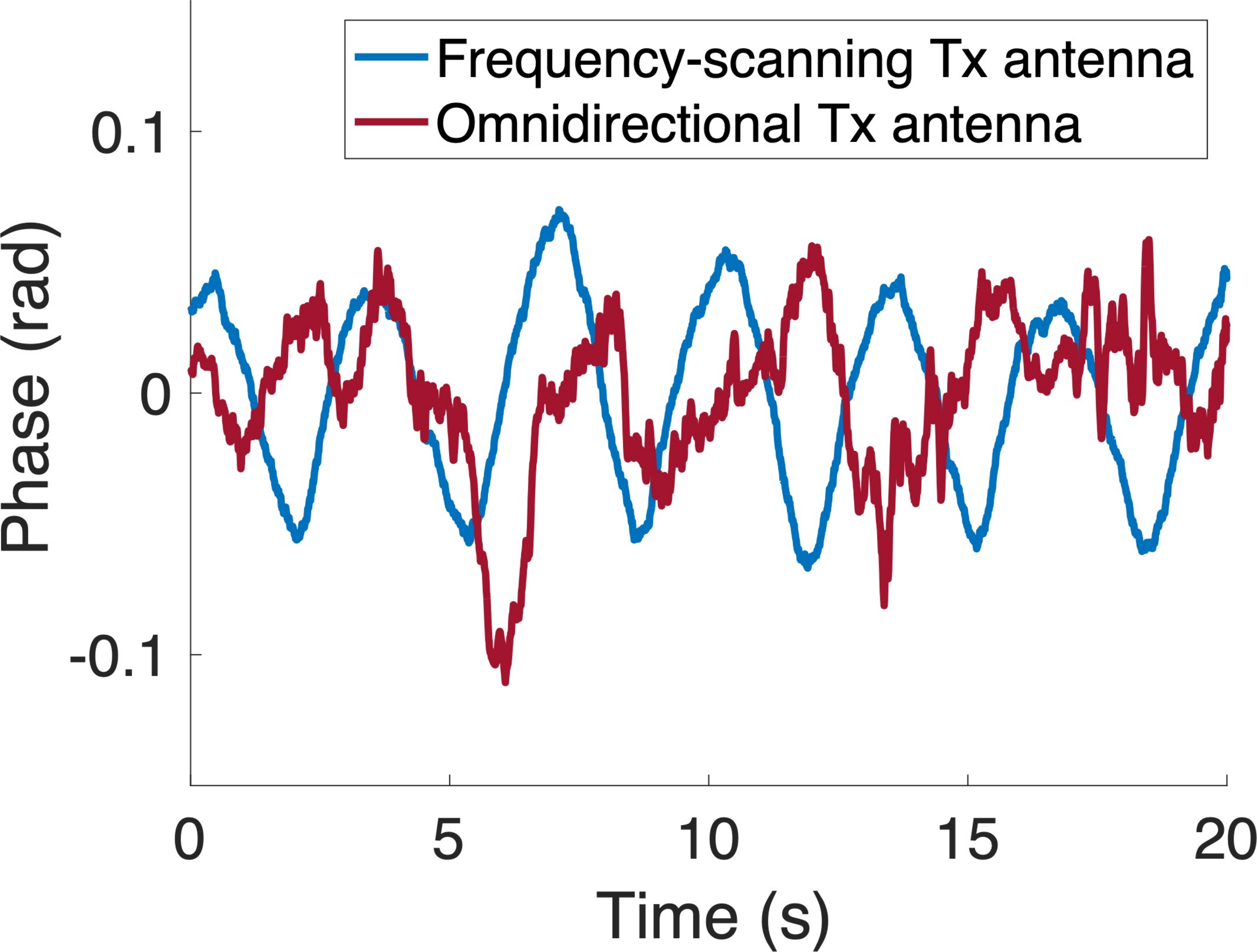}
        \label{fig5_6b}
        \hspace{0in}
    }
    \vspace{-0em}
    \caption{Evaluation on interference-resistant multi-target in a living room.}
    \vspace{-0em}
    \label{fig5_56}
\end{figure*}

\begin{figure*}[t]
    \centering
        \subfloat[A cluttered office with one target and a rotating fan]{
        \includegraphics[height=1.55in]{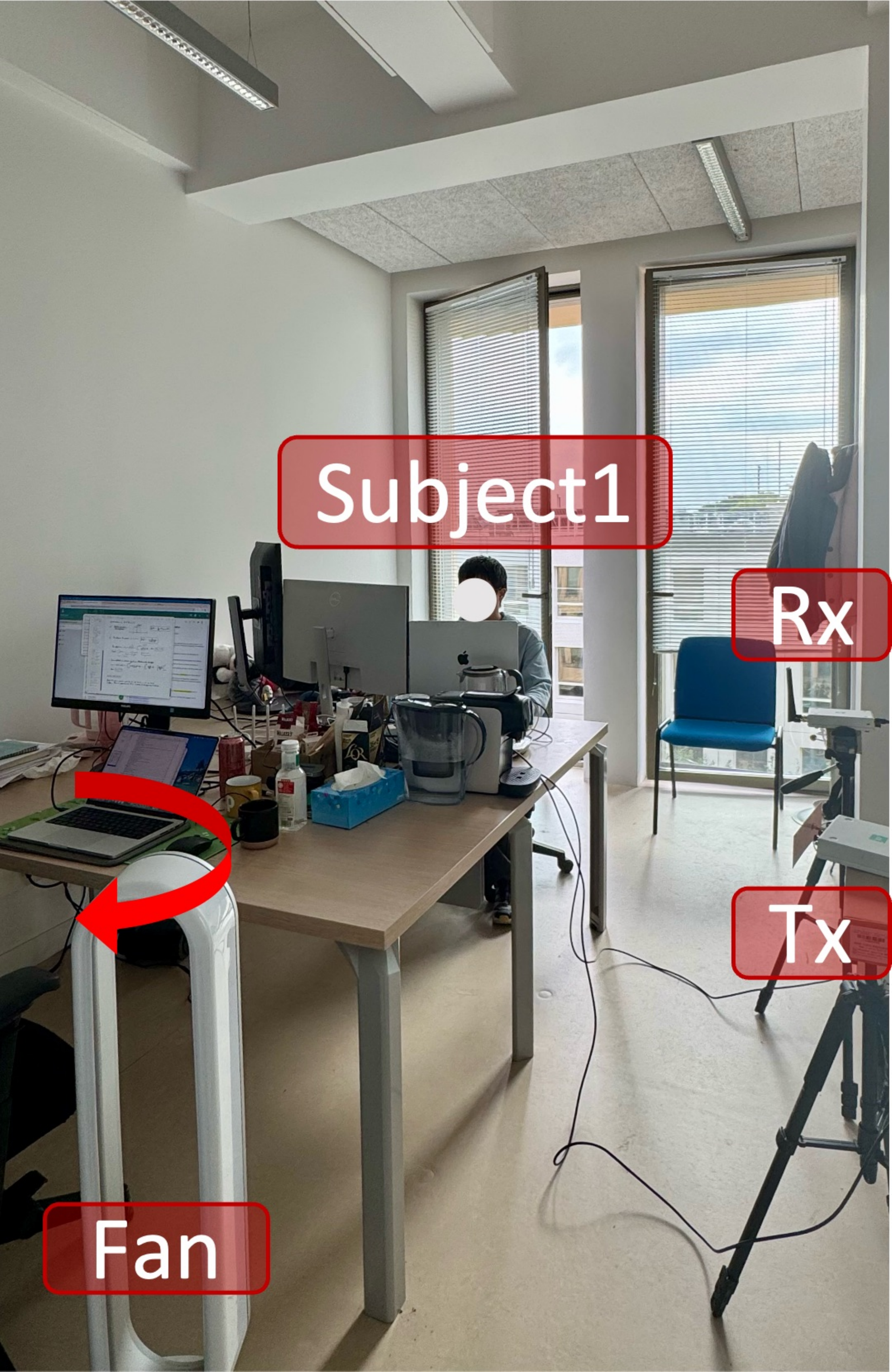}
        \label{fig5_8a}
    }
        \hspace{0.1in}
        \subfloat[An office with two subjects and a fan close to one subject]{
        \includegraphics[height=1.55in]{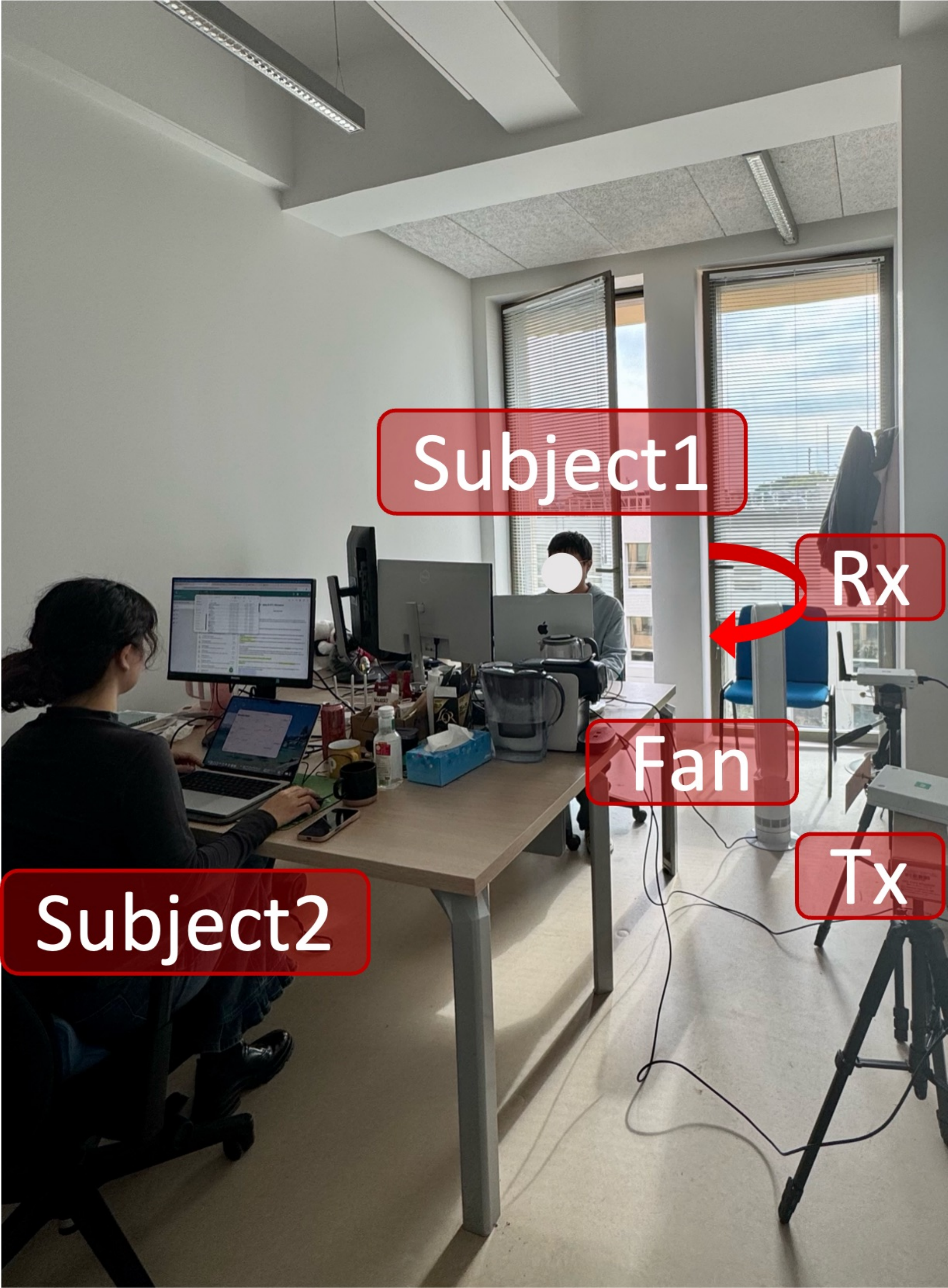}
        \label{fig5_8b}
    }
        \hspace{0.1in}
        \subfloat[The impact of rotating fan on signals transmitted by an omnidirectional antenna and a FSA]{
        \includegraphics[height=1.55in]{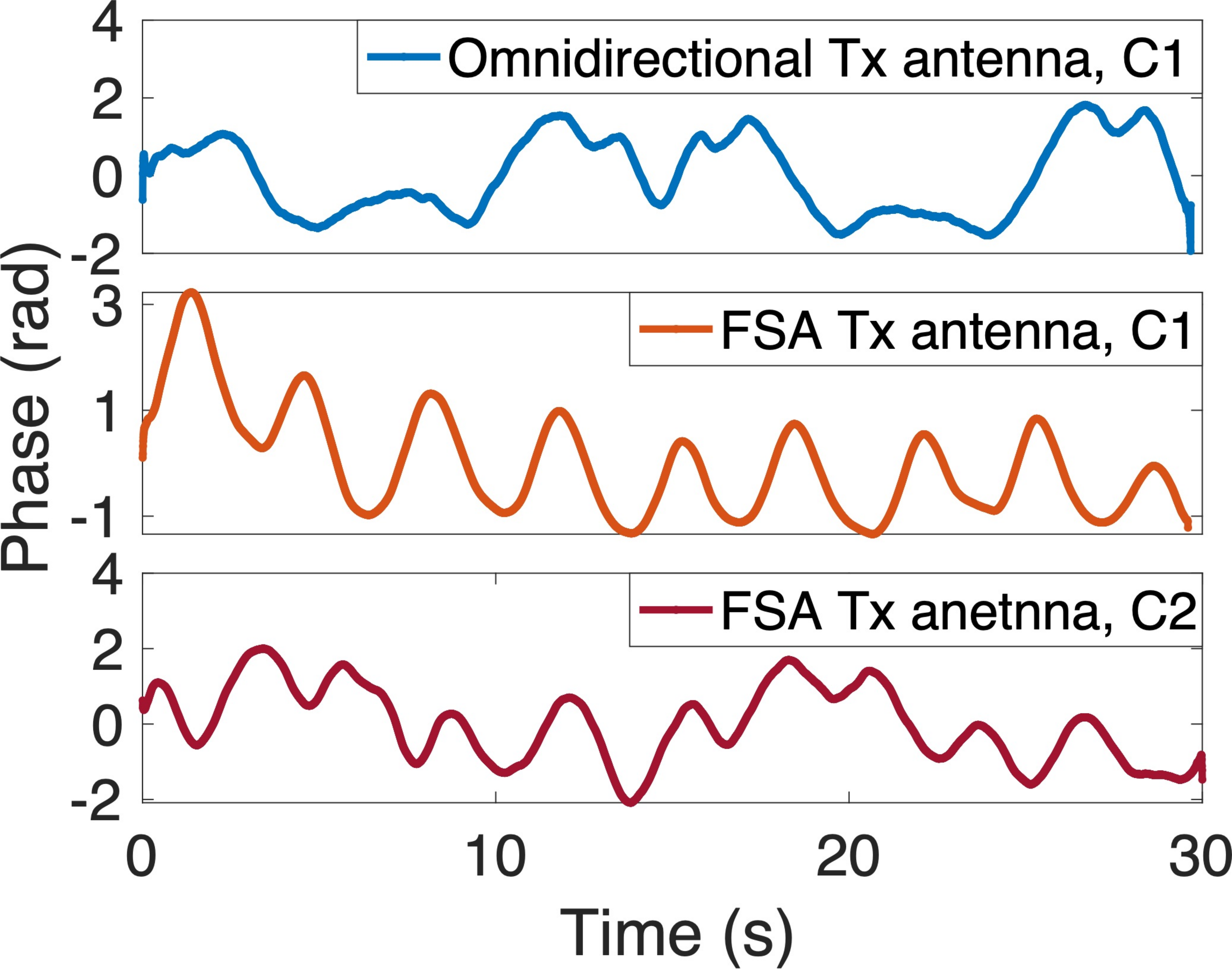}
        \label{fig5_8c}
    }
        \hspace{0.1in}
    \subfloat[Respiration rate estimation error]{
        \includegraphics[height=1.55in]{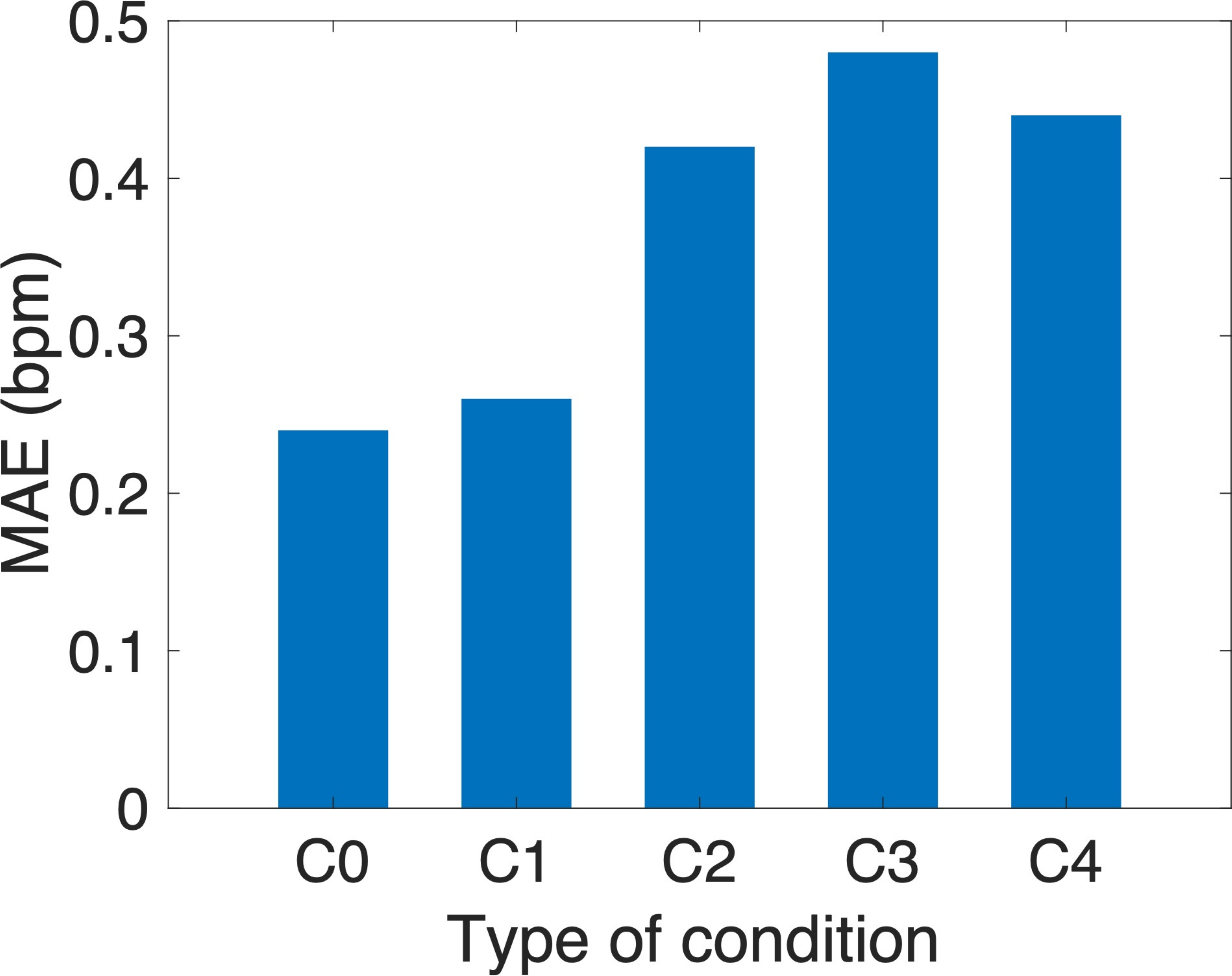}
        \label{fig5_8d}
        \hspace{0in}
    }
    \vspace{-0em}
    \caption{Respiration monitoring in a cluttered office.}
    \label{fig5_8}
\end{figure*}

As illustrated in Figure~\ref{fig5_2}, we conduct experiments in three scenarios, including two smart building environments and one smart home setting. In the smart building use cases, the Wi-Fi transceiver pair, with the FSA as the transmitter, is deployed at key locations such as a hallway corner (Figure~\ref{fig5_2a}) and an entry point (Figure~\ref{fig5_2b}) to support intrusion detection. These applications require not only detecting whether a person crosses a virtual boundary but also determining the direction from which the person is approaching. However, the state-of-the-art solution can only detect the occurrence of crossing events without identifying the movement direction~\cite{wang2024understanding}. In contrast, \systemname{} leverages angle estimation to assign each walking path to one of two angular regions, thereby enabling accurate classification of walking directions using only a single antenna. To evaluate this capability, we design four representative walking trajectories as shown in Figure~\ref{fig5_3a}.  As shown in Figure~\ref{fig5_3b}, the resulting confusion matrix for the four walking trajectories in the two environments demonstrates that \systemname{} achieves 100\% classification accuracy, confirming its effectiveness in direction-aware intrusion detection.

We further explore the use of FSA-based sensing in a smart home scenario (Figure~\ref{fig5_2c}), where different directions relative to the FSA correspond to doorways leading to separate rooms. By estimating the direction of movement, the system can infer which room the user is entering, enabling personalized automation and context-aware services. To assess this function, we conduct experiments in which a user walks from Room R0 to one of three destination rooms. The user’s initial position in R0 is not fixed, introducing variability and better reflecting real-world behavior. Throughout the movement, the system continuously estimates the direction. By detecting the final angle before signal variations disappear (as the sensing area in R0 is limited), we can infer the room the user has entered. To evaluate the robustness of this approach, we test four different receiver placements, as illustrated in Figure~\ref{fig5_3c}. Across all configurations, \systemname{} achieves a minimum room classification accuracy of 87\%, as shown in Figure~\ref{fig5_3d}, demonstrating its capability to support room-level localization for smart home applications.

\vspace{-0em}

\subsection{Case Study 2: Interference-Resistant Multi-Target Respiration Monitoring}
In this section, we demonstrate the potential applications of FSAs for respiration monitoring. Due to their ability to disperse Wi-Fi signals across different directions, FSAs are particularly well-suited for monitoring multiple targets and exhibit robustness against interference from moving objects. To validate this capability, we conduct experiments in three typical indoor environments, including a living room, a bedroom, and an office. These settings are selected to comprehensively evaluate the system’s resilience against both complex static multipath effects and dynamic motion interference.

We begin with experiments conducted in a smart home scenario, as illustrated in Figure~\ref{fig5_2c}. The experiment setup is shown in Figure~\ref{fig5_5a}, where the FSA is positioned facing a sofa to enable simultaneous respiration monitoring for one to three seated individuals. We extract the subcarriers corresponding to three distinct directions. As shown in Figure~\ref{fig5_5b}, the respiration rate estimation error remains consistently below 0.62 bpm regardless of the number of monitored targets. These results confirm the high accuracy and scalability of the proposed system in multi-target respiration monitoring tasks. Next, we evaluate the effectiveness of mitigating interference. In this experiment, we monitor two seated individuals while introducing a disturbance by having one subject stand up and walk away from the sofa, as shown in Figure~\ref{fig5_6a}. Figure~\ref{fig5_6b} compares the received signals under two antenna configurations. When using a standard omnidirectional antenna at the transmitter, the received signal becomes dominated by high-frequency fluctuations induced by the moving subject. As a result, the respiration waveform is obscured and becomes difficult to recover. In contrast, when the FSA is employed, the respiration signal remains clearly visible. By extracting the subcarrier associated with the target’s direction, the system effectively suppresses the interference from unrelated motion in other directions.

We further evaluate our system in more challenging environments to validate its robustness against severe multipath effects caused by clutter and dynamic interference introduced by various moving objects. Specifically, we begin with a cluttered office space measuring 4.6 m $\times$ 2.7 m, where two individuals sit at a shared desk, as shown in Figures~\ref{fig5_8a} and~\ref{fig5_8b}. The two positions correspond to approximately 30$\degree$ and 30$\degree$ within the FSA’s FoV. To evaluate system robustness under different interference conditions, we first examine the baseline case C0, where only one person is seated at either of the two positions. In conditions C1 and C2, we introduce an operating fan that rotates periodically. In C1, the fan is placed near the transmitter as shown in Figure~\ref{fig5_8a}, while in C2, it is placed close to subject 1 and also closer to the transceivers than the target subject, creating stronger multipath and dynamic interference. We compare the received signals using both an omnidirectional antenna and the FSA. As shown in Figure~\ref{fig5_8c}, with the omnidirectional antenna, the respiration waveform is overwhelmed by low-frequency fluctuations caused by the fan’s rotation. In contrast, with the FSA, we extract the subcarrier aligned with the target’s direction. The respiration waveform remains clearly observable, even under the strong interference in condition C2.

To further increase realism, we introduce a second person into the office. In condition C3, the second subject performs natural office activities such as typing, drinking water, and adjusting posture, while subject 1 remains still and breathes normally. The fan continues running and remains closer to subject 1. Finally, in condition C4, both subjects remain static and breathe normally, while the fan continues to run as the sole source of interference. We summarize the respiration monitoring performance under all five conditions using the mean absolute error metric, as shown in Figure~\ref{fig5_8d}. The results confirm that the system consistently maintains low estimation error across different subjects, positions, and interference scenarios. These findings demonstrate the robustness, reliability, and scalability of FSA-based sensing in real indoor environments with multiple sources of interference.

\begin{figure}[t]
        \centering
        \vspace{-0em}
        %\vspace{-1em}
        \subfloat[A bedroom with multiple moving objects]{
        \includegraphics[height=1.3in]{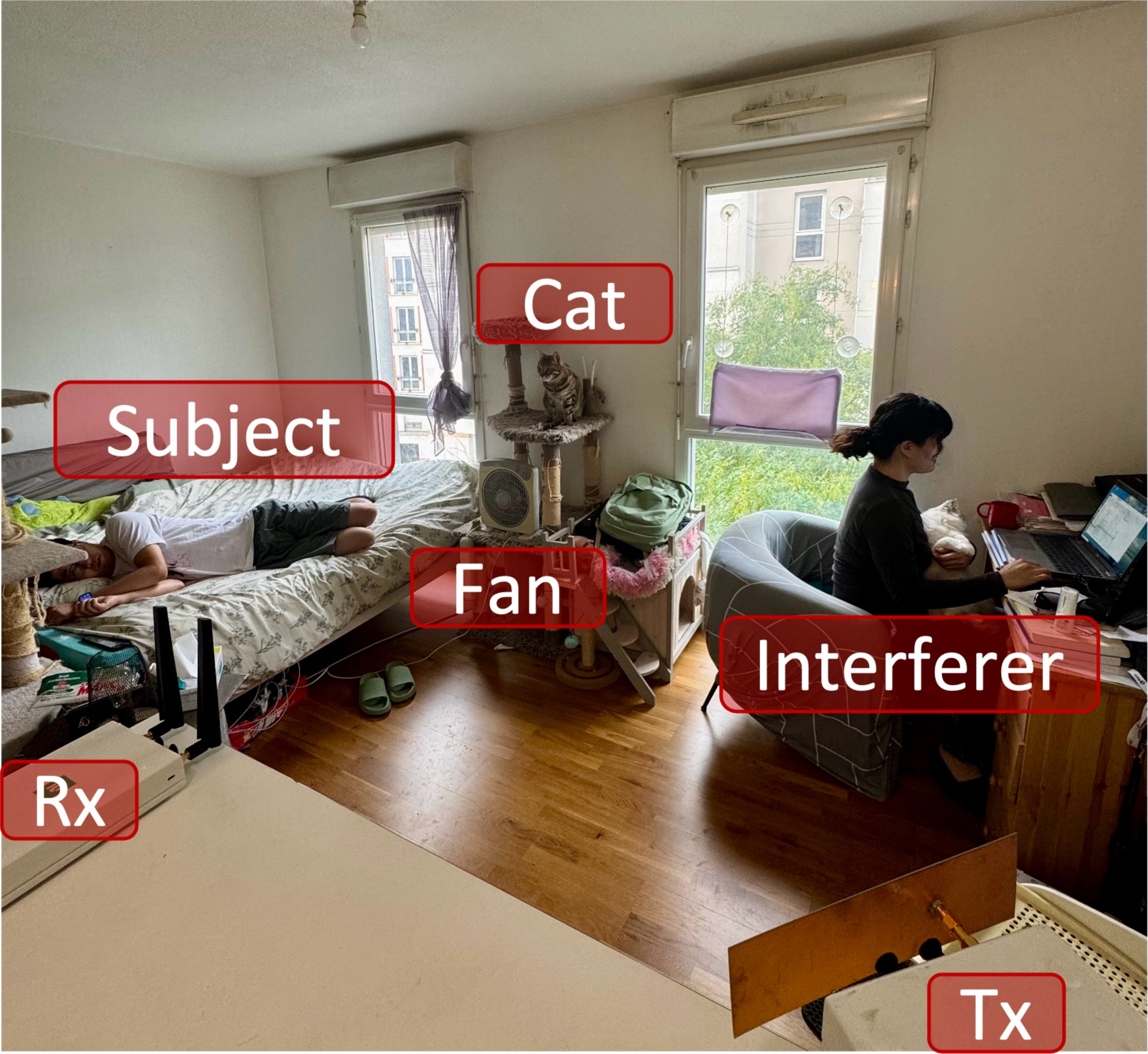}
            \label{fig5_7a}
        }
        \hspace{0.0in}
        \subfloat[Respiration rate estimation error]{
        \includegraphics[height=1.3in]{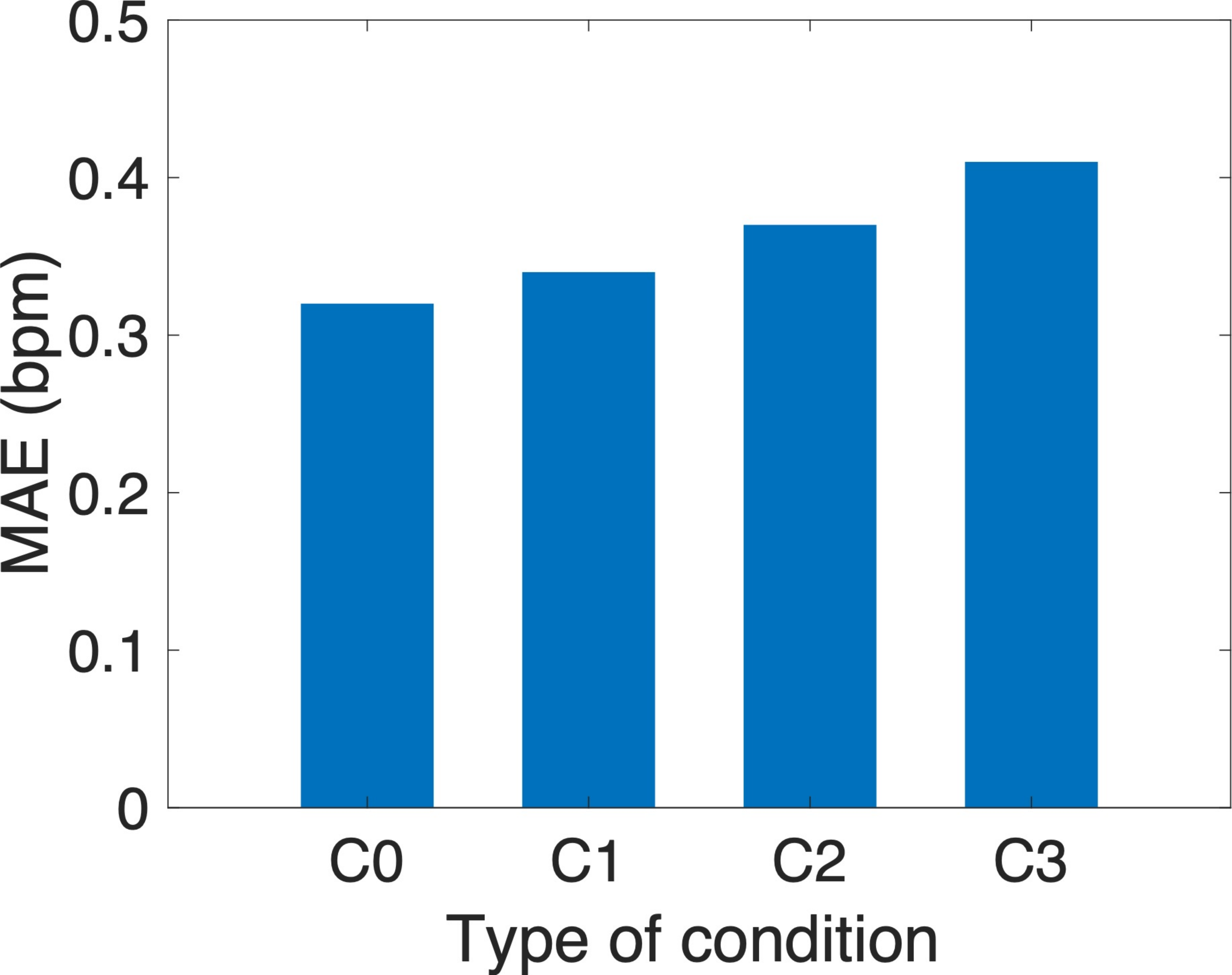}
            \label{fig5_7b}
        }
        \vspace{-0.0em}
        \caption{Respiration monitoring with several moving objects in a bedroom. }
        \label{fig5_7}
        %\vspace{-0.15in}
\end{figure}

We also deploy the system in a real-world bedroom environment, as shown in Figure~\ref{fig5_7a}. The room measures 4.9 m $\times$ 4.2 m. The sensing target lies on a bed positioned at approximately 20$\degree$ within the FSA’s FoV. The surrounding environment includes cluttered furniture and multiple sources of motion-induced interference. We evaluate the system under four progressively more challenging interference conditions. In the baseline condition C0, only the target remains on the bed. In condition C1, a fan is placed near the bed, introducing periodic mechanical motion. In condition C2, a cat is introduced and moves freely on the cat tower, producing irregular and unpredictable motion. Finally, condition C3 adds a human interferer working at a desk, introducing natural motion patterns such as typing or shifting posture. For each condition, we extract the subcarrier corresponding to the bed’s direction to obtain the respiration waveform. Since the interfering sources are located in different directions, their influence is significantly attenuated. Each experiment lasts 30 seconds and is repeated five times. As shown in Figure~\ref{fig5_7b}, even under the most challenging condition (C3), the system accurately estimates the target’s respiration rate with a low error of 0.41 bpm.

\section{Discussion and Limitation}

\subsection{Compatibility with Communication}

While our proposed FSA design enables direction-aware sensing, it operates independently of the main Wi-Fi communication system. Specifically, we envision deploying the FSA as a dedicated sensing antenna, while conventional MIMO architectures continue to support communication. This separation of sensing and communication not only simplifies system integration but also avoids the resource competition in MIMO-based ISAC systems. By isolating sensing from communication, our design offers a lightweight and infrastructure-compatible approach, highlighting its potential in future Wi-Fi sensing and ISAC deployments.

\subsection{Using FSAs as Receiving Antennas}
FSA can also function effectively as a receiving antenna, meaning that the frequency-scanning property allows signals from different spatial directions to be mapped to distinct subcarriers, effectively making the FSA operate as a directional antenna with frequency-dependent beam patterns. By analyzing the received signals across subcarriers, the receiver can directly estimate the angle of incoming signals. This capability further enhances the versatility and application potential of our design.

\subsection{Antenna Limitations and Future Extensions}
Our current FSA achieves a 60$\degree$ FoV within a 160 MHz Wi-Fi band and demonstrates clear frequency–angle mapping. While effective, the beam distribution across frequencies is not perfectly uniform, and some directions exhibit stronger gains as shown in Figure~\ref{fig4_1}. Future work may explore optimizing the resonator layout to improve the uniformity of beam direction and gain distribution. Furthermore, extending the antenna design to the 6 GHz Wi‑Fi band is a promising direction for future research, as the latest protocol supports 320 MHz channels that could enable a larger FoV. Moreover, our design can further be adapted to other wireless sensing modalities, such as utilizing 4G/5G, RFID, LoRa, UWB, and mmWave, offering opportunities for broader sensing applications.

\subsection{Application Scenarios}
The proposed system design has the potential to be integrated into other Wi-Fi sensing applications for estimating target angles, such as gesture recognition, activity recognition, and localization. By leveraging angle estimation, Wi-Fi sensing systems can gain additional spatial awareness, improving recognition accuracy and enhancing user interactions. Beyond Wi-Fi sensing applications, our antenna design also has the potential to be extended to a wide range of angle-related applications. For instance, it can be utilized for eliminating device or body motion by extracting reflection signals from objects or body areas at different angles~\cite{chang2024msense,chang2024mmecare}. It can also contribute to imaging and point cloud generation~\cite{zhao2018through,regmi2021squigglemilli,qian20203d,prabhakara2023high,lai2024enabling} by mapping signal reflections from different angles, enabling spatial reconstruction using various wireless signals. 
\section{Related Work}

\subsection{Applications of FSAs}

Typically, FSAs are utilized for signal source direction finding. Some studies have focused on THz communication networks for link discovery~\cite{ghasempour2020single}, THz device localization~\cite{kludze2022quasi}, and security purposes~\cite{yeh2020security}. FSAs are also investigated for estimating the directions of RFID tags with respect to an RFID reader~\cite{martinez2011frequency,gil2021frequency,gil2022direction}. This principle can also be used for Bluetooth device direction estimation~\cite{poveda2020frequency}. More recently, BIFROST~\cite{sun2023bifrost} proposes an FSA-based Wi-Fi backscatter design, which can be used for Wi-Fi device localization in non-LoS conditions. However, existing FSA-based solutions primarily focus on estimating the direction of an active transmitter rather than passive dynamic reflections from human motion. In this paper, we focus on utilizing FSAs for passive Wi-Fi sensing and address specific challenges for sensing in aspects of both antenna design and signal processing.

\subsection{Direction-Aware Wi-Fi Sensing}

Wi-Fi sensing has become an increasingly popular research area, leveraging signal reflections to infer human motion and activities. Recent advancements have incorporated angle estimation techniques to provide richer spatial awareness~\cite{kotaru2015spotfi,soltanaghaei2021tagfi,ren2022gopose,wu2023enabling,xie2024robust}. Wi-Vi~\cite{adib2013see} investigates the use of an antenna array with the MUSIC algorithm for through-wall human motion detection. Phasebeat~\cite{wang2017phasebeat} leverages multi-antenna for multi-target respiration monitoring. IndoTrack~\cite{li2017indotrack} proposes a device-free indoor human tracking system utilizing Doppler speed and AoA joint estimation. Multiple pairs of Wi-Fi transceivers with antenna arrays are required for walking trajectory recovery. MD-Track~\cite{xie2019md} further involves ToF estimation together with angle information to improve tracking accuracy. Wincent~\cite{ren2021winect} presents a 3D human pose tracking system for free-form activity recognition using an antenna array. Regardless of the method, existing Wi-Fi sensing systems must use multiple antennas to obtain target angle information. In contrast, \systemname{} can achieve direction-aware Wi-Fi sensing using a single antenna, reducing hardware cost and complexity.

\section{Conclusion}

In this paper, we propose a novel integrated antenna and algorithm design that enables direction-awareness in Wi-Fi sensing applications using a single antenna. The core innovations of WiRainbow lie in two complementary components: (1) a coupled-resonator-based antenna architecture that significantly enhances the antenna’s dispersion property, overcoming the narrow FoV limitation inherent in traditional traveling-wave FSA designs; and (2) a robust SSNR-based algorithm framework specifically designed to distinguish human-induced signal variations from multipath reflections. Such a design can be adaptable across sensing tasks of different motion scales and speeds. We implement and evaluate the prototype system, demonstrating that WiRainbow achieves reliable angle estimation and interference-resistant multi-target sensing under realistic multipath conditions, making it a scalable and cost-effective solution for diverse Wi-Fi sensing applications.

\section*{Acknowledgments}
This work is supported by the European Union through the Horizon EIC pathfinder challenge project SUSTAIN (No. 101071179), the Innovative Medicines Initiative 2 Joint Undertaking project IDEA-FAST (No. 853981), the project of Shenzhen Science and Technology Innovation Committee (KJZD20240903103300002 and KCXFZ20240903094011015), and the project of Scientific Research Platform and Projects of Colleges and Universities of Guangdong Province Department of Education (No. 2023ZDZX3082).

\bibliographystyle{ACM-Reference-Format}
\balance
\bibliography{sample-base}

% \appendix

\end{document}